\documentclass[english,aps, superscriptaddress,twocolumn]{revtex4}
\usepackage[T1]{fontenc}
\usepackage[latin9]{inputenc}
\usepackage{textcomp}
\usepackage{relsize}
\usepackage{amsmath}
\usepackage{graphicx}
\usepackage{amssymb}
\makeatletter

\DeclareRobustCommand{\lyxmathsym}[1]{\ifmmode\begingroup\def\b@ld{bold}
  \def\rmorbf##1{\ifx\math@version\b@ld\textbf{##1}\else\textrm{##1}\fi}
  \mathchoice{\hbox{\rmorbf{#1}}}{\hbox{\rmorbf{#1}}}
  {\hbox{\smaller[2]\rmorbf{#1}}}{\hbox{\smaller[3]\rmorbf{#1}}}
  \endgroup\else#1\fi}

\usepackage{upgreek}

\usepackage{float}
\usepackage{babel}

\begin{document}

\preprint{This line only printed with preprint option}

\title{Reorganization of columnar architecture in the growing visual cortex}

\author{Wolfgang Keil}

\email{wolfgang@nld.ds.mpg.de}

\affiliation{Max-Planck Institute for Dynamics and Self-Organization, 37073 G\"ottingen,
Germany}
\affiliation{Lewis Sigler Institute for Integrative Genomics, Princeton University,
08544 Princeton, NJ}
\author{Karl-Friedrich Schmidt}

\affiliation{Institute of General Zoology and Animal Physiology, Friedrich-Schiller
University, 07743 Jena, Germany}

\author{Siegrid L\"owel}

\affiliation{Institute of General Zoology and Animal Physiology, Friedrich-Schiller
University, 07743 Jena, Germany}

\author{Matthias Kaschube}

\email{kaschube@princeton.edu}

\affiliation{Lewis Sigler Institute for Integrative Genomics, Princeton University,
08544 Princeton, NJ}
\begin{abstract}
Many cortical areas increase in size considerably during postnatal
development, progressively displacing neuronal cell bodies from each
other. At present, little is known about how cortical growth affects
the development of neuronal circuits. Here, in acute and chronic experiments, we study
the layout of ocular dominance (OD) columns in cat primary visual
cortex (V1) during a period of substantial postnatal growth.
We find that despite a considerable size increase of V1, the spacing between columns is
largely preserved. In contrast, their spatial arrangement changes systematically over this period. 
While in young animals columns  are more band-like, layouts become more isotropic in mature animals. 
We propose a novel mechanism of growth-induced reorganization that is
based on the `zigzag instability', a dynamical instability observed in several
inanimate pattern forming systems.
We argue that this mechanism is inherent to a wide class of
models for the activity-dependent formation of OD columns.
Analyzing one member of this class, the Elastic Network model, we show that this mechanism
can account for the preservation of column spacing and the specific mode of reorganization of OD
columns that we observe. We conclude that neurons systematically shift their selectivities
during normal development and that this reorganization is induced by the  cortical expansion during growth.
Our work suggests that cortical circuits remain plastic for an extended period in development in order to facilitate the modification of neuronal circuits to adjust for cortical growth.
\end{abstract}
\maketitle

\section{Introduction}
The brain of most mammalian species grows substantially during postnatal
development without a significant change in the number of neurons. 
The human brain, for instance, weighs on average 350g in newborns and 1400g in adult males \cite{dekaban_78}. In cat,
the neocortical volume increases from $\approx$1000mm\textsuperscript{3} at birth
to $\approx$4500mm\textsuperscript{3} in adulthood \cite{villabanca_00}. 
Consistently, the surface area of cat primary visual cortex (V1) increases postnatally
by a factor of 2.5 between week 1 and week 12 \cite{Duffy:1998p2219}. 
This size increase implies that the
distance between any two neuronal cell bodies grows on average by
a factor of 1.6 during postnatal development. 
Do neuronal processes such as axons and dendrites simply elongate or 
are larger changes needed to accommodate this growth? 
It is not known at present how the brain achieves permanent adjustment of its functional wiring   
to the changing physical proportions while at the same time being fully functional at every moment. 
While the importance of mechanical factors is appreciated in a number of growth related phenomena in biology such as morphogenesis \cite{adam:10}, heart development \cite{vermot:09}  and tumor growth \cite{kumar:09}, their possible impact on functional aspects of neural circuits has received relatively little attention.

In cat primary visual cortex (V1), much of the growth takes place during a period in which most parts of
the visual field are already represented  \cite{Sireteanu:1982p4332} and many neurons
have already reached fairly mature levels of selectivity. For instance,
the selective response of visual cortical neurons to inputs from one
eye or the other, called ocular dominance (OD), can already be visualized
at postnatal week 2 in cat V1 \cite{Crair:2001p1095} (Fig. \ref{fig:area_increase}A). 
OD is organized into columns which can be labeled over the full extent of V1 as early as
week 3 \cite{Rathjen:2003p2212} (Fig. \ref{fig:area_increase}B). 
For these properties, this system is well suited for studying the impact of cortical 
growth on neural circuitry. 

What happens to cortical columns when the cortex is growing in size? 
The seemingly simplest scenario, in which new columns are inserted 
into the cortex, appears rather implausible since 
the number of neurons remains largely constant during this period \cite{Cragg:1975p481}.
In fact, most of the area increase is due to the generation of glial cells,
the addition of more vasculature and connective tissue, and the myelinization
of axons. To a lesser extent it also reflects the outgrowth and elaboration
of axonal and dendritic processes \cite{Purves:1994p3921}. Therefore, 
a different scenario has been suggested, sometimes referred to as the `balloon effect', 
in which  columns expand by a similar factor as the surrounding
cortical tissue \cite{Duffy:1998p2219}. 
In this study we start out by testing the balloon hypothesis for the case of OD columns in 
cat visual cortex. We show that the expected expansion of columns during 
cortical growth does not take place. Instead, columnar layouts reorganize over the considered period 
and become more isotropic in older animals.  
These observations strongly argue against a simple balloon-like expansion and imply that 
cortical circuits can respond to the constraints arising during growth by a different as yet unknown mechanism. 

In order to account for our empirical observations, a fraction of neurons must either shift their relative spatial location  or, alternatively, alter their functional response properties. Although appealing, the former possibility is difficult to address at present, since little is known about coherent motion of groups of neurons in response to mechanical tension \cite{essen:97,smith:09,anava:09}. In contrast, a large body of experimental and theoretical work exists addressing phenomena related to cortical plasticity and demonstrating the impressive susceptibility of neural circuits  to changes in activity patterns, most often in the context of OD  \cite{katz_02,Hensch:2005p590}. 
Furthermore, it is noteworthy that in the two most intensely studied animal models for cortical plasticity, namely the cat and the mouse, the period of brain and body growth coincide with and end at about the same time as the period that allows for intense restructuring of neuronal connections \cite{daw:92, harlan:96,lehmann:08,gall:68}. 
In this study, we therefore explore the latter possibility and analyze the predicted reorganization in models for the activity-dependent formation of OD columns.  Based on general properties of these models, we develop a novel scenario of growth-induced cortical reorganization. Characteristic features of this reorganization as well as the time scale on which it evolves are in good agreement with the changes in columnar layout we observe during postnatal growth in cat V1.
\section{Results}
\subsection{The spacing of OD columns is preserved
over a period of cortical growth}

\begin{figure*}
\includegraphics[width=12cm]{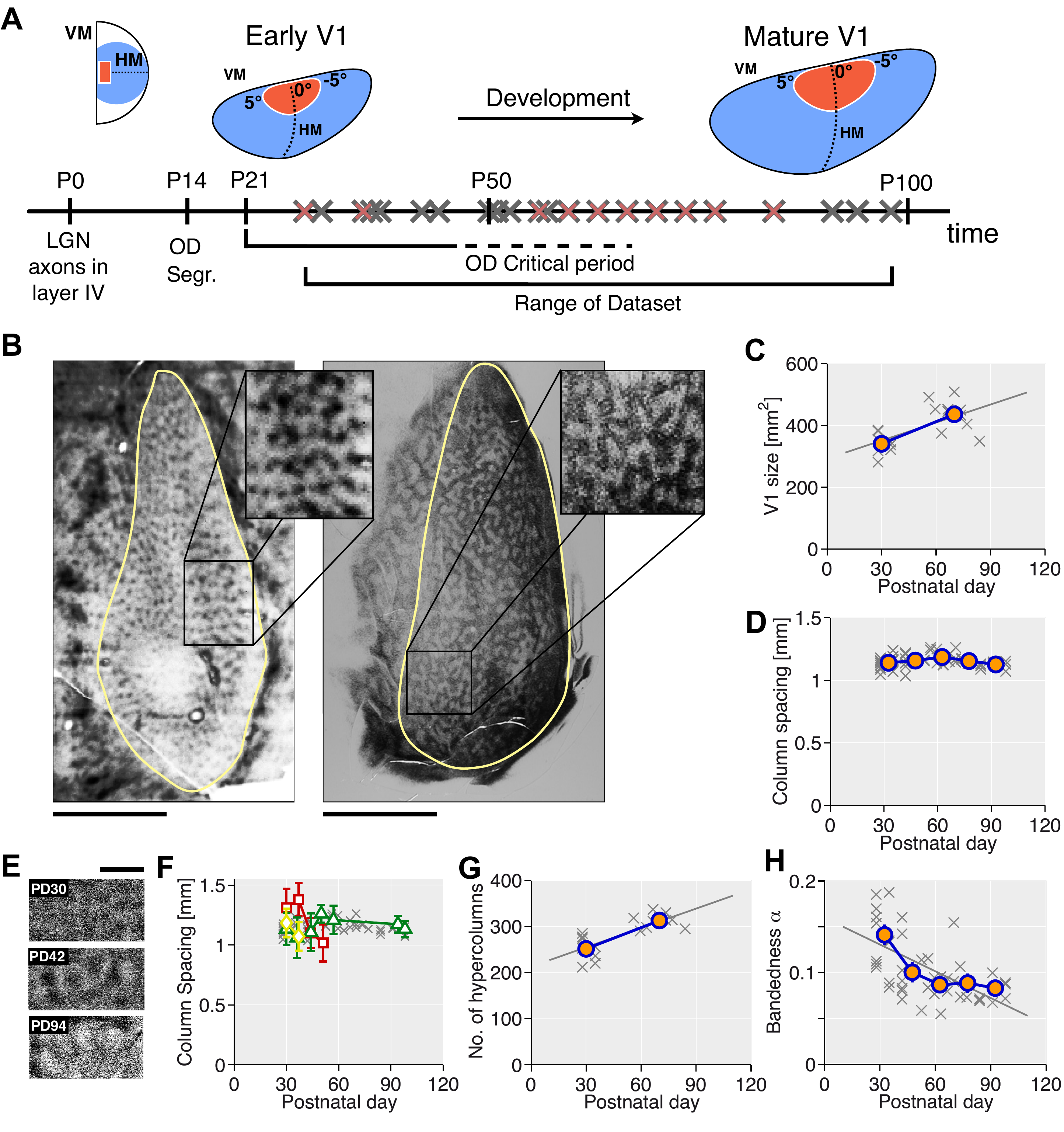}
\caption{Reorganization of OD columns in cat visual
cortex (V1) over a period of cortical growth.
({A}) Growth of cat V1, the representation of the
visual hemi-field (schematic; blue area, V1; red area, central visual field representation; HM, horizontal meridian; VM, vertical
meridian), and the time line of cat OD development  \cite{Crair:2001p1095}.
Our dataset includes 2-DG/proline-labeled hemispheres from kittens
between PD28 and PD98 (2-DG, N=37; proline, N=4) and 
chronic optical recordings between PD30 and PD98 (crosses mark individual data points).
 ({B}) Two representative examples of OD layouts. Left, at PD30 (2-DG); right, at PD60 (proline).
 Yellow lines mark V1 borders (scale bar, 10mm). 
({C}) Area sizes of V1 for the N=18 kittens with complete reconstructions of V1 (red-marked crosses in A). Blue-orange dots represent averages over pools of sizes denoted by the gray numbers (error bars smaller than symbol size). Gray crosses mark values for individual animals. Gray line shows linear regression ($r=0.6248$, $p<0.006$). 
(D) OD Column spacings $\Lambda$ do not increase over this period  ($r=-0.034$, $p<0.825$). 
(E) OD columns in cat V1 by intrinsic signal optical imaging (same animal; high-pass filtered, \textit{SI Appendix}; scale bar, 1.5mm).  
(F) Column spacings $\Lambda$  for the case in (E) (green triangles) and for two other cases (red boxes, yellow diamonds) corroborating the results in (D) (shown in light gray for comparison;  error bars by bootstrapping). 
(G-H) While the  number of hypercolumns $N_{HC}$   increases over this period  ({G})  ($r=0.7720$, $p<0.0002$),  
 the bandedness $\alpha$ decreases considerably  (H)  ($r=-0.5754$, $p<6\cdot10^{-5}$). 
\label{fig:area_increase}}
\end{figure*}

We first measured the size increase of cat primary visual cortex (V1)
during early postnatal development (Fig. \ref{fig:area_increase}A).
We labeled complete layouts of
OD columns in V1 visualized by either 2-{[}$^{14}$C]-deoxyglucose (2-DG)  or {[}$^{3}$H]-proline in kittens at
different ages between postnatal day (PD) 28 and PD98 ({N=18} hemispheres, Fig. \ref{fig:area_increase}B). 
 V1 was readily discernible by its distinctive columnar activation
pattern in comparison to the labeling in surrounding cortical areas \cite{Kaschube:2003p430}. 
In particular, V1 was distinguished from V2 based 
on its considerably smaller column spacing. 
We observed a size increase
of about a factor of 1.3 between two groups centered at PD30 and PD70 ($r=0.6248$,
$p<0.006$) (Fig. \ref{fig:area_increase}C). To reduce 
possible influences of genetic variability \cite{Kaschube:2002p429},
we analyzed a littermate couple on PD30 and PD72. Consistent with our previous results, V1 area was a factor of 1.46 larger in the older kitten. 
Thus, our analyses confirmed previous 
studies \cite{Duffy:1998p2219,Rathjen:2003p2212} 
by observing a considerable size increase of V1
during cat postnatal development. 

We next asked whether the spacing of OD columns increases by a corresponding
factor over this period. 
First, we measured the  column spacing $\Lambda$ of 
2-DG/proline-labeled OD patterns in {N=41} hemispheres 
between PD28 and PD98 (data includes the N=18 hemispheres used for the analyses of V1 sizes.
To obtain accurate estimations of column spacings, we used the wavelet-method  
introduced in \cite{Kaschube:2002p429,Kaschube:2003p430}  
(\textit{SI Appendix}). As shown in Fig. \ref{fig:area_increase}D,
 column spacings varied between 1.05mm and 1.28mm, but did not
show a significant increase over this period ($r=-0.034$, $p<0.825$).
Consistent with this observation, the column spacings of the two
littermates differed by less than 10\%, despite their difference in V1 size 
of 46\%.

To follow the development column spacings in individuals
we visualized OD columns by chronic optical imaging 
({N=3} hemispheres, total age range: PD30-PD98) 
(Fig. \ref{fig:area_increase}E).  We quantified their spacings by the above wavelet-method (Fig. \ref{fig:area_increase}F).
While column spacings based on optical recordings exhibited larger variability	 compared to the 2DG/proline data (Fig. \ref{fig:area_increase}F), we found no systematic increase of column spacings in individual animals, thus confirming the conclusions drawn from the 2DG/proline data. Increased variability might be explained by the substantial interareal variability of OD column spacings \cite{Kaschube:2003p430} together with the fact that the imaged regions were much smaller than V1 and may have shifted with age. 

Taken together, both the 2-DG/proline
data and the chronic optical recordings demonstrate that the postnatal
growth of cat V1 is not accompanied by a corresponding
increase in the spacing of OD columns, strongly arguing against the balloon scenario. 
\subsection{OD columns reorganize during cortical growth}
An increase of area without a change in column spacing indicates an increase in the
number of hypercolumns. The concept of a hypercolumn is related to that of a 
functional module and denotes a cortical unit  containing a full set of values for any given 
set of receptive field parameters \cite{horton:05}.
We roughly estimated the typical size of a hypercolumn by $\Lambda^2$  (\textit{SI Appendix}) 
and define the number of hypercolumns in a map by  
 $N_{HC}=A/\Lambda^{2}$  \cite{Kaschube:2003p430}, where $A$ is its total area.
Fig. \ref{fig:area_increase}G shows that  for the N=18 completely reconstructed hemispheres from Fig. \ref{fig:area_increase}C 
the number of hypercolumns  $N_{HC}$ increases significantly  ($r=0.7720$, $p<0.0002$). At PD28, V1 contains on average $260\pm40$  hypercolumns ({N = 5}), increasing to $319\pm17$ ({N=6}) at PD72 (increase of 23\%). 

To reveal more directly the reorganization of OD columns, we analyzed  
a third parameter called bandedness $\alpha$ that characterizes the structural properties of local pattern elements 
(\cite{Kaschube:2002p429,Kaschube:2003p430},
 Fig. \ref{fig:wavelet_sheme}). Large values of $\alpha$ indicate layouts composed of regular stripe-like parallel domains,
whereas small values indicate more isotropic layouts such as bended stripes or patches.
Such quantitative  evaluation of the spatial organization of columns was possible in N=39 hemispheres. 
We found that the bandedness $\alpha$ decreases by almost a factor of 2
 from an average of $0.14\pm0.02$ (N=13) at PD35 to an average of $0.0832\pm0.006$ at PD95 (N=5)
($r=-0.5754$, $p<6\cdot10^{-5}$; Fig. \ref{fig:area_increase}H). 
This systematic decrease in bandedness indicates that OD columns, while largely preserving their initial spacing, reorganize
and develop more isotropic layouts over time. 
\subsection{Modeling OD column formation with growth}
\begin{figure*}
\includegraphics[width=8.5cm]{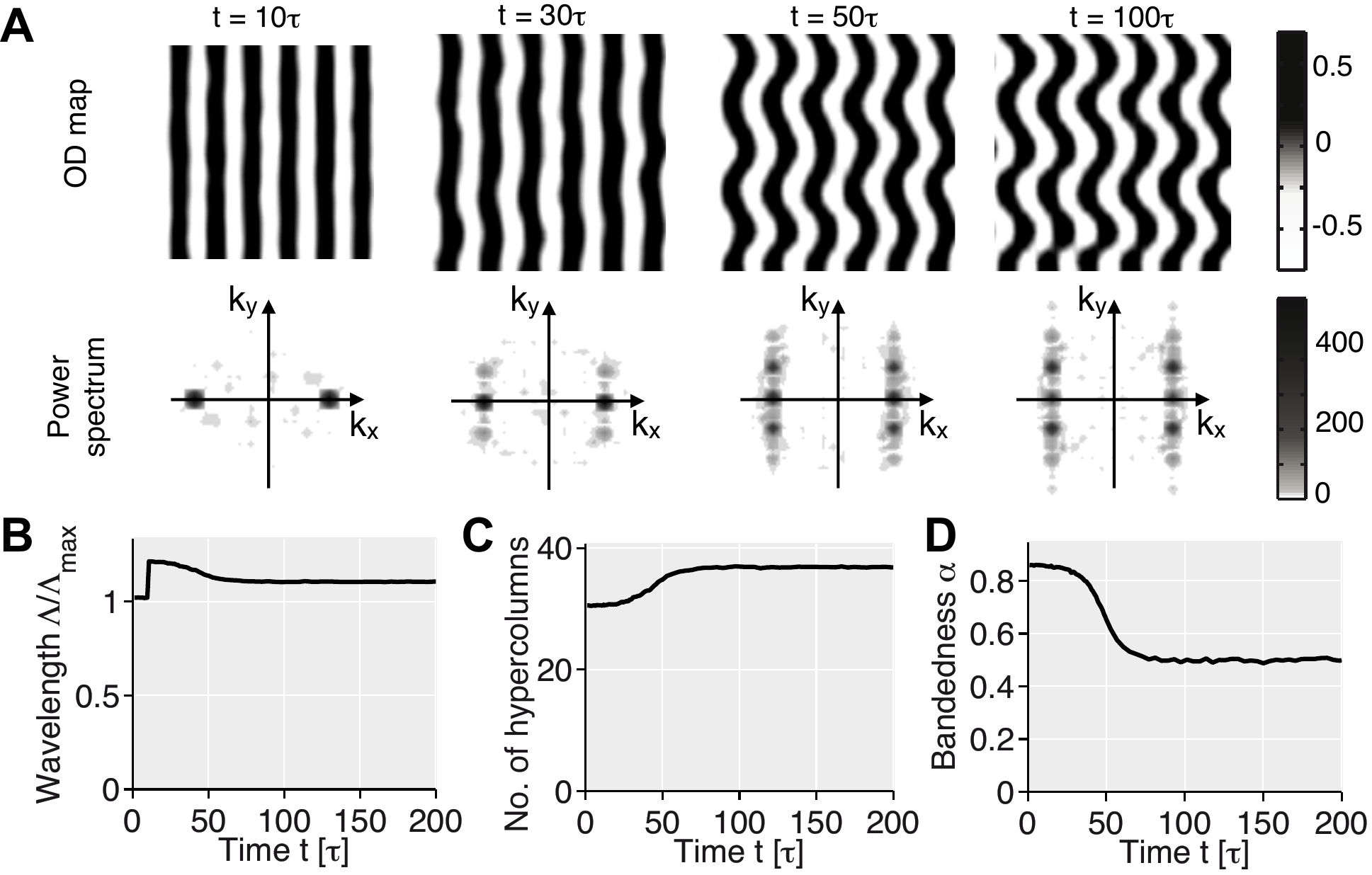}
\caption{Expansion-induced reorganization in models for OD formation. 
({A}) Snapshots of a simulation starting from a near
steady state solution of the EN model \cite{Durbin:1990p1196}, 
a stripe-like OD pattern (upper row) corresponding to a single Fourier mode in the 
power spectrum (lower row) ($\eta=0.025,\, r=0.15$).
After instantaneous area increase (linear extent by a factor of 1.18, i.e. $\delta k/k_{\textnormal{max}}=-0.15$, at 10$\tau$), OD domains bend sinusoidally and additional Fourier modes appear at $(k_{\textnormal{max}}+\delta k)\mathbf{\vec{x}}\pm q_{y}\mathbf{\vec{y}}$. ({B-D})  This reorganization is captured by the
 column spacing $\Lambda$ ({B}), the number of hypercolumns $N_{HC}$ ({C}),
and the bandedness $\alpha$ ({D}) (time in units of  the time scale $\tau$  of OD segregation).
\label{fig:zigzag_simulated} }
\end{figure*}
To understand these experimental observations, we studied  cortical growth in models for the activity-dependent self-organization of OD columns. For specificity, we focussed on the well-studied 
Elastic Network (EN) model \cite{Durbin:1990p1196,Wolf:1998p1199,Goodhill:2000p2141} (Material and Methods, \textit{SI Appendix}). 
Solutions in the absence of growth are shown in Fig. \ref{fig:elastic_net_model} and Movie SV1. Linear stability analysis around the initially nonselective cortex (\textit{SI Appendix}, \cite{Wolf:1998p1199}) identifies a control parameter $r$ describing the distance from pattern onset. A pattern of OD columns forms for $r>0$. The analysis also defines an intrinsic timescale $\tau=1/r$, on which the segregation of columns takes place, and a spatial scale $\Lambda_{\textnormal{max}}$ that is roughly equal to the  column spacing of the developing OD pattern. As in other models for the self-organization of  OD columns \cite{Swindale:1980p1946,miller_89,Obermayer:1992p1200,Goodhill:2000p2141}, this spatial scale arises from the effective recurrent interactions which have a 'Mexican-hat' structure
(local facilitation, nonlocal suppression). 
In agreement with previous work \cite{Wolf:1998p1199,Goodhill:2000p2141}, we find in simulations  that  an OD pattern emerges after a few $\tau$  (Fig. \ref{fig:elastic_net_model}, Movie SV1). The only steady state solutions we observe are parallel OD stripes.  

As a simple way to mimic cortical growth, we started from steady state solutions and abruptly increased isotropically the size of the simulated system without changing the other model parameters (i.e without increasing the width of the Mexican-hat; Materials and Methods and \textit{SI Appendix}). 
Fig. \ref{fig:zigzag_simulated} displays snapshots of a typical example
of such a simulation  (see also Movie SV2). Upon size
increase at $t=10\tau$, stripes start to bend sinusoidally (Fig.
\ref{fig:zigzag_simulated}A, upper row). In the power spectrum, this corresponds to the growth of
two additional Fourier modes with wave vectors $\sim(k_{\textnormal{max}}+\delta k)\mathbf{\vec{x}}\pm q_{y}\mathbf{\vec{y}}$ (Fig. \ref{fig:zigzag_simulated}A, lower row). 

We quantitatively analyzed this reorganization by the wavelet-method used above. 
The column spacing $\Lambda$ increases abruptly  at 10$\tau$,  but subsequently decreases to close its initial value
(Fig. \ref{fig:zigzag_simulated}B).  The number of hypercolumns $N_{HC}$ increases persistently (Fig. \ref{fig:zigzag_simulated}C),  
while the bandedness $\alpha$   decreases significantly over this period  (Fig. \ref{fig:zigzag_simulated}D).
Thus, the growth-induced bending of OD columns largely restores  the 
initial spacing and results in a bandedness  drop similar to what we observe in experiment (Fig. \ref{fig:area_increase}).  
\subsection{A general mechanism of growth-induced reorganization}
\begin{figure*}
\includegraphics[width=7.5cm]{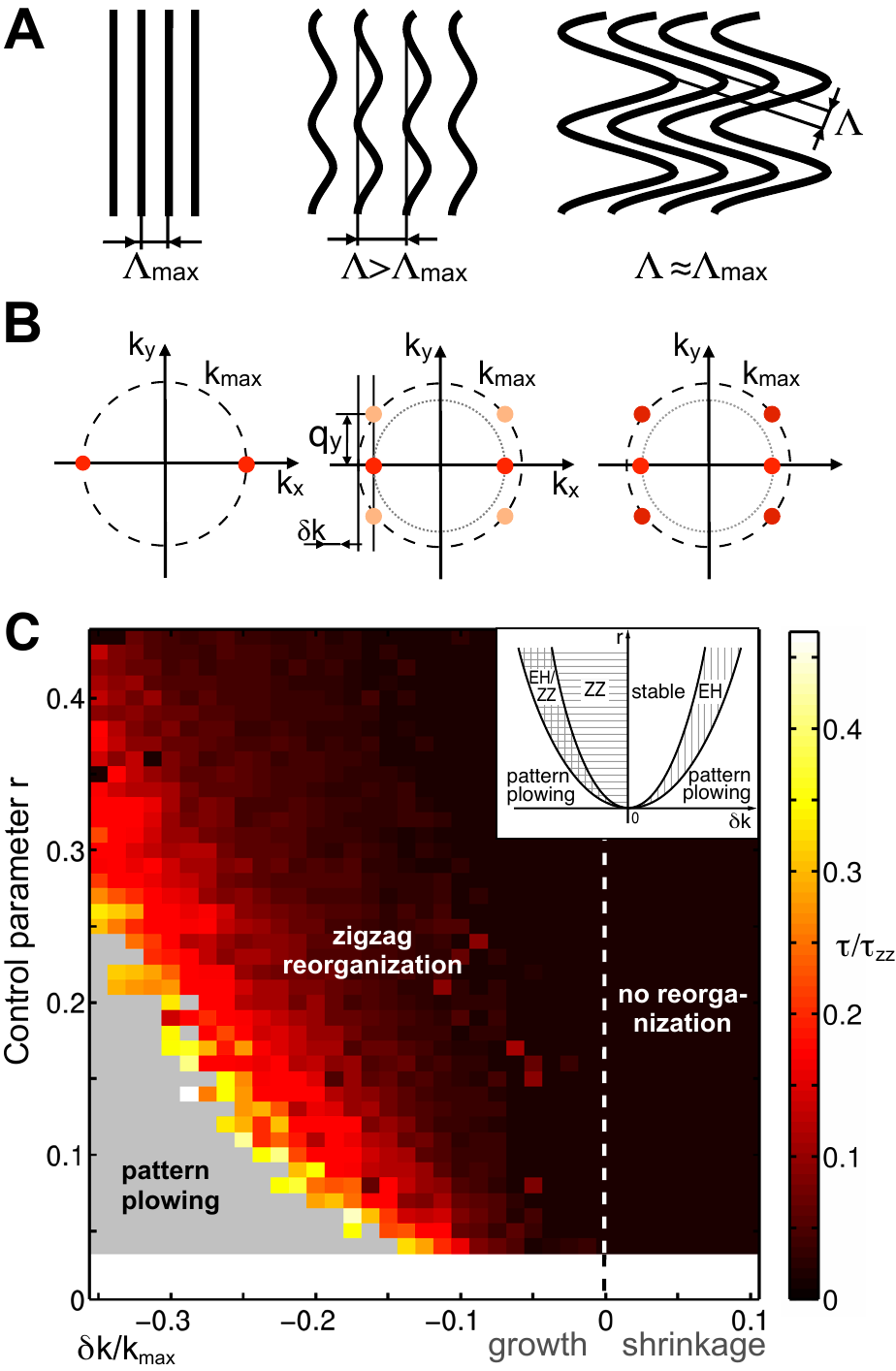}
\caption{The ZZ instability provides a general framework for understanding growth-induced cortical 
reorganization. 
(A) The ZZ mechanism: Initially, OD stripes exhibit the spacing $\Lambda_{\textnormal{max}}$ (left). 
Upon expansion (middle), OD stripes recover their initial spacing locally through bending of OD stripes (right). 
(B) In Fourier space: Two new Fourier modes grow at
$(k_{\textnormal{max }}+\delta k)\mathbf{\vec{x}}\pm q_{y}\mathbf{\vec{y}}$ restoring the initial
wave number  $ k_{\textnormal{max}}$ of the pattern. 
(C) Regime and time scale of ZZ instability in the EN model. Instantaneous isotropic expansion (regime left of white dashed line) induces a ZZ reorganization that evolves on a timescale $\tau_{ZZ}\gg \tau$ (\textit{SI Appendix}).  
For very large expansion, a complete new pattern forms (gray region).
For area decrease (right of white dashed line) no ZZ instability was observed.
No simulations were carried out in the white region, since the simulation
time diverges for $r\rightarrow0$ (\textit{SI Appendix}). 
Inset: Predicted regime of ZZ instability for 2-dimensional, relaxational, isotropic dynamics close to pattern onset ($r\ll1$) (redrawn from \cite{cross_greenside_09}; horizontally striped region,  ZZ instability;  vertically striped region,  Eckhaus (EH) instability, i.e. insertion/elimination of a stripe, see \textit{SI Appendix}).  
\label{fig:busse_balloon_EN} }
\end{figure*}

We show in the following that this type of expansion-induced reorganization of OD columns  
is caused by a zigzag (ZZ) instability \cite{newell_69},  a type of dynamical instability which has been widely studied in the theory of pattern formation \cite{Cross:1993p922,cross_greenside_09}. Fig. \ref{fig:busse_balloon_EN}, {A and B}, illustrate the ZZ mechanism.
This instability is typical for the wide class of relaxational, rotationally symmetric models in which a two-dimensional pattern forms by a finite wavelength instability \cite{Cross:1993p922}. This class includes the EN model, as we outline in the \textit{SI Appendix}, and many other OD models (e. g. \cite{Swindale:1980p1946,miller_89,Obermayer:1992p1200,Goodhill:2000p2141}).

A theory  \cite{Cross:1993p922,cross_greenside_09} for this model class exists predicting the regime of the  ZZ instability (inset in Fig. \ref{fig:busse_balloon_EN}C).
However, strictly speaking, this theory is valid only in a narrow parameter region close to the point of pattern onset at $r=0$. 
We therefore analyzed numerically the behavior of the EN model further away from onset by  probing systematically a large set of instantaneous size increases  and testing for growing ZZ modes (Fig. \ref{estimate_tau_zz}, \textit{SI Appendix}). 
We observed that the regime of ZZ instability is very large (Fig. \ref{fig:busse_balloon_EN}C). Similar to the theoretical predictions  \cite{Cross:1993p922,cross_greenside_09},  even a slight expansion  results in a ZZ instability and its regime increase parabolically with the control parameter $r$. 
Moreover, the induced reorganization evolves on a time scale $\tau_{ZZ}$ that can be more than an order of magnitude larger than the time scale $\tau$ of OD column segregation.
 \subsection{Realistic Growth Scenarios}
\begin{figure*}
\includegraphics[width=13cm]{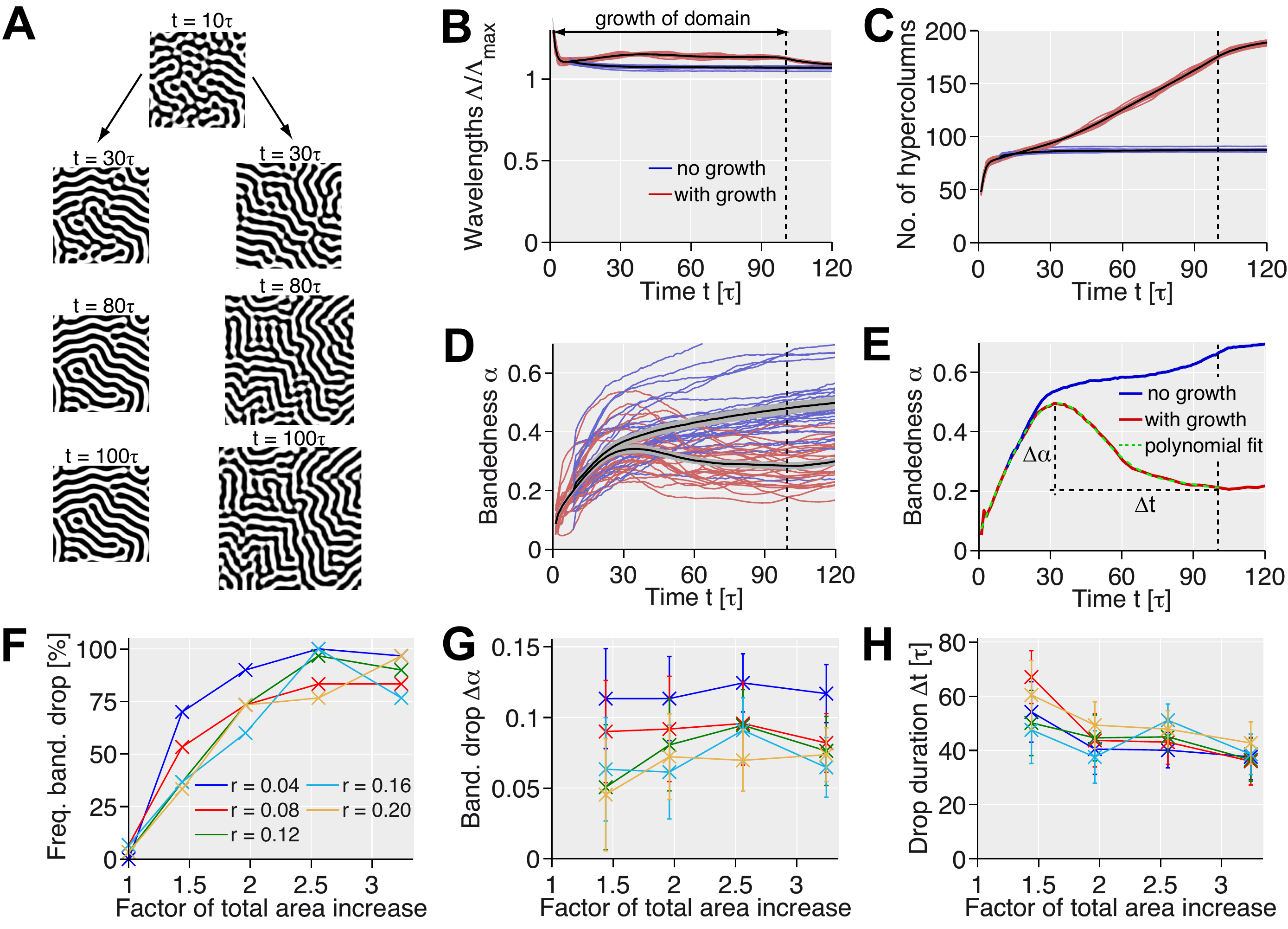}
\caption{ZZ-type reorganization of OD columns in a realistic  growth scenario.
({A}) Snapshots of EN model simulations with isotropic linear area increase by a factor of 2.56 between $t=0$ and $t=100\tau$
(left column) and without size increase after $t=10\tau$ (right column)
($r=0.16,\,\eta=0.025$). ({B-D}) Time courses of  column
spacing $\Lambda$ ({B}), number of hypercolumns $N_{HC}$ ({C}),
and  bandedness $\alpha$ ({D}) for 30 pairs of simulations as in A. Black curves are averages, gray regions s.e.m.. Despite the large area increase, growing systems display an only mild and transient increase in $\Lambda$  implying a strong increase in $N_{HC}$, while $\alpha$   typically drops considerably. ({E}) Quantification of bandedness drop by its strength $\Delta\alpha$ and duration  $\Delta t$. ({F-H}) The percentage of simulations showing a substantial bandedness drop (defined by  $\Delta\alpha>0.05$, $\Delta t>15\tau$) (F), the average size  $\Delta\alpha$ of such drops ({G}) and their average duration $\Delta t$ ({H}) evaluated for various total area increases and control parameters $r$ (N=50 simulations per data point, error bars indicate s.e.m.). 
\label{fig:realistic_growth_sim} }
\end{figure*}
Finally, we show that growth-induced reorganization
shows signatures of the ZZ instability
even if the initial OD layout is not a simple stripe pattern and the increase in 
system size follows a continuous growth scenario.   
To approximate realistic conditions, we initiated our simulations with
the nonselective state and linearly
increased the linear extent of the simulated regions by 
a factor of 1.6 (factor of 2.56 in area increase) between $t=0\tau$ and $t=100\tau$. 
Fig. \ref{fig:realistic_growth_sim}A shows that layouts appear to be more bended in a ZZ-fashion when
compared to simulations for which we stopped growth after $t=10\tau$ .
Typically, the column spacing $\Lambda$ in growing systems increases only transiently (Fig. \ref{fig:realistic_growth_sim}B) implying that the hypercolumn number $N_{HC}$ increases persistently (Fig. \ref{fig:realistic_growth_sim}C). 
The bandedness $\alpha$ is relatively variable across solutions  
reflecting the large diversity of the evolving OD layouts (Fig. \ref{fig:realistic_growth_sim}D and Fig. \ref{fig:elastic_net_model}). 
However, whereas in virtually all simulations without growth $\alpha$  increases nearly 
monotonously  (Fig. \ref{fig:elastic_net_model} and Movie SV1), in growing systems $\alpha$ typically dropped considerably 
 reflecting the ZZ-type reorganization of OD columns (Fig. \ref{fig:realistic_growth_sim}A and Movie SV3). 

We systematically studied the growth-induced reorganization by varying the control parameter $r$ and testing different area  
increases (factor of 1, 1.44, 1.96, 2.56 and 3.24, Fig. \ref{fig:realistic_growth_sim}, E to G; \textit{SI Appendix}). 
We measured the difference $\Delta\alpha$ between the first maximum in bandedness  and 
the subsequent minimum and the time interval $\Delta t$ 
between these two bandedness extrema
(Fig. \ref{fig:realistic_growth_sim}E; based on 8th-order polynomial least square fit, \textit{SI Appendix}).
Whereas  drops with $\Delta\alpha>0.05$ and $\Delta t>15\tau$ occurred in only
$<6\%$ of the  non-growing systems, already with moderate growth, they are present in a large fraction of systems 
(Fig. \ref{fig:realistic_growth_sim}F).  
For smaller values of $r$ the drop is generally more pronounced. 
Both the average size $\Delta\alpha$ and the  duration
$\Delta t$  depend only weakly on the total area
increase and drops last on average more than $40\tau$ (Fig. \ref{fig:realistic_growth_sim}G and H).

Thus, also for more realistic growth scenarios, the induced reorganization
exhibits key features of a ZZ instability, in particular the only mild and transient increase in column spacing 
and the prominent and long lasting drop in bandedness.  
Intriguingly, these features also describe  the mode of reorganization we observe in experiment (Fig. \ref{fig:area_increase}).  
Moreover, if we  identify the model time unit 
$\tau$ with  roughly one day in cat postnatal development - an assumption which may be justified by experiments showing that OD columns segregate within a few days (e.g. \cite{Crair:2001p1095}) -   
we observe that even the time scales on which these changes evolve, agree fairly well between model and experiment. 
This suggests that the reorganization we observe in experiments is caused by cortical expansion through a mechanism that is based on the ZZ instability.
%
%
%
%
%
%
%
%
%
%
%
%
\section{Discussion}
\subsection{Cortical expansion, the absence of the balloon effect and the range of the 'Mexican-hat'}
$\phantom{\,}$Our empirical data provides evidence against the so-called `balloon effect' \cite{Duffy:1998p2219} by showing that OD columns do not simply  expand during cortical growth, but largely maintain their spacing. At first sight, a balloon-like expansion may seem plausible. For instance, \textit{in-vitro} connected neurons, when moderately pulled  apart, readily extend their axonal arbors to prevent disruption, thereby achieving  neurite growth rates of up to 1mm/day \cite{smith:09}. However, mechanical tension on one axonal branch can strongly influence the arborization of other branches of the same neuron \cite{anava:09}, indicating that expansion-induced responses can be rich and may lead to nontrivial collective behavior in expanding networks of interconnected neurons. 

Models for the activity-dependent formation of OD columns can reproduce the absence of the balloon effect if the width of the lateral interactions is kept fixed  during growth as we assume in this study. In the EN model, an effective intracortical interaction of Mexican-Hat-type (see Fig. \ref{fig:elastic_net_model}A) arises from the interplay between the coactivation of cortical regions and a tendency for neighboring neurons to acquire similar response properties \cite{Durbin:1990p1196}. Even if the interaction range increases by only half the rate of the cortex, we observe a ZZ-type reorganization accompanied by a bandedness drop  (\textit{SI Appendix}, Fig. S8). However, in this case also the column spacing increases systematically. Thus, these models can be reconciled with our data only if the interaction range does increase only little during growth.

There are several possibilities of why the range of effective lateral interactions might not increase during growth. It is conceivable that the width of interaction could depend on the lateral spread of dendritic arbors. Limits on an increase of the arbor size during growth might be imposed by a tendency of neuronal circuits to minimize the total length of wiring (see e.g.  \cite{chklovskii:02}) and could be achieved by synaptic pruning \cite{Cragg:1975p481}.
Alternatively, interactions of Mexican-Hat type  could  arise if the time scales  of  the dominant  inhibitory synapses are small compared to excitatory synapses \cite{kang:03}. Thus, a possible shift from a dominance of smaller towards larger synaptic timescales during the period of growth could partly compensate for the increase of the distance between neurons. Finally,  Mexican-hat type interactions could arise by excitatory connections that at larger distances preferentially target inhibitory interneurons. In this case, the width of interactions may depend on the strength of inhibition \cite{Hensch:2005p590} which, appropriately adjusted, could keep the range of the Mexican-hat constant during growth.  

\subsection{Reorganization vs. Displacement} 
$\phantom{\,}$In the scenario of growth-induced reorganization proposed in this paper, increasing distances between cell bodies alter the effective lateral interactions between neurons, thereby inducing shifts in the response properties in a fraction of them. Alternatively, one may explore a scenario in which neuronal response properties are preserved, and the mature columnar layout is obtained by an  inhomogeneous  displacement of cells. Strong intra-columnar connections may provide the necessary mechanical stability for keeping cells within columns closer to one another. However, the ability of neurons to rapidly extend their axonal arbors in response to mechanical tension \cite{smith:09} raises doubts about expansion-induced forces being strong enough to promote   inhomogeneous displacement of cells.  
Following the spatial positions as well as functional response properties of many cells experimentally, e.g. by chronic 2-photon microscopy, could help to disentangle these two hypotheses.
Moreover, experiments that, instead of altering the patterns of neural activity, probe the cortex mechanically during development may reveal novel insights about the role of mechanical tension in cortical development.
A better understanding of the interplay between tension-mediated and activity-driven mechanisms in shaping neural circuits \textit{in-vivo}   could shed new light on normal development and cortical growth, but potentially  also on the response of neural network function to cortical lesions and brain tumors. 

The two scenarios have different implications for the hypercolumn as a functional cortical unit. A shift in OD in individual neurons would alter the set of stimulus representations in a hypercolumn. It would be interesting to monitor simultaneously other neuronal selectivities and test whether they co-develop in a systematic fashion, e.g. by improving coverage uniformity \cite{swindale:00} over time. 
On the other hand, a pure displacement of groups of neurons would distort the original hypercolumn and result in systematic inhomogeneities in the cortical representation of the visual field position. Such inhomogeneities may be detectable in the mature cortex even without the necessity of technically very challenging chronic experiments.
\subsection{Relation to Previous Work}
$\phantom{\,}$A longitudinal optical imaging study \cite{Muller:2000p1269} of OD columns in a single strabismic cat reported
an increase of OD column spacing between PD27 and PD61 consistent with the slight but not significant increase we observe over this period (Fig. \ref{fig:area_increase}D).
A chronic imaging study in ferret reported a fairly stable spatial organization of orientation columns between PD30 and PD55  \cite{chapman:96}.  However, a more recent study \cite{Kaschube:2009} analyzing orientation columns in the cat between PD 35 and PD105  observed changes in local column spacing that  were coordinated between V1 and V2. Consistent with the present study, the average spacings in V1 and V2  remained largely constant over this period.   
A theoretical study \cite{Oster:2006p1955} of a one-dimensional model of OD development during cortical growth predicts 
a splitting of OD stripes analogous to the Eckhaus instability (\textit{SI Appendix}). 
As we show here, two-dimensional models exhibit a much richer dynamics and behave qualitatively different. 
\subsection{A Novel Function of Plasticity in Normal Development}
$\phantom{\,}$The impressive ability of cortical circuits to reorganize during and after the critical period has been demonstrated in numerous studies by artificially manipulating cortical activity, e.g. by monocular deprivation (see \cite{Hensch:2005p590} for a review). 
However, relatively little is known at present  about the role of cortical plasticity for normal cortical development \cite{Crair:2001p1095,katz_02}, (but see \cite{Kaschube:2009,wang:10} ). As we point out in this study, the period of cortical plasticity in cat visual cortex overlaps with the period of postnatal cortical growth  \cite{daw:92, harlan:96}. 
While the peak  of the classical critical period is around PD30 \cite{olson:80},  cortical plasticity does not cease after the critical period, but rather declines gradually \cite{Hensch:2005p590}. 
It is readily conceivable that this plasticity may be exploited by the cortex in order to accommodate for growth-induced changes.  
Interestingly, the reorganization we report here is largest close to the peak of the critical period. (Fig. \ref{fig:area_increase}H).
Furthermore, key features of this reorganization are reproduced by modeling OD formation as self-organization based on cortical plasticity. Thus, we conclude that  cortical plasticity may play an important role in normal development through facilitating growth-related modifications of neuronal circuits.

\section{methods\label{sec:methods}}
subsection{Experiment}
OD patterns were labeled with 2-{[}$^{14}$C]-deoxyglucose
(2-DG) autoradiography  after monocular stimulation
of the animals or by {[}$^{3}$H]-proline autoradiography after injection
of the labeled proline into one eye which labels the thalamocortical
afferents of that eye in cortical layer IV (see \cite{Kaschube:2003p430} and therein).
OD columns were recorded by intrinsic signal optical imaging following \cite{schmidt_06}. 

\subsection{Model}
OD is described by a real valued
field $o(\mathbf{x},t)$, where $\mathbf{x}$ represents the position
on the cortical surface and $t$ time. Negative/positive values of $o(\mathbf{x},t)$ indicate
a preference for inputs from the ipsilateral/contralateral eye. The dynamics of this field is given by
\begin{equation}
\partial_{t}o(\mathbf{x},t)=\left\langle \left[s_{o}-o(\mathbf{x},t)\right]A_{\sigma}(\mathbf{x}, \mathbf{S},o(\cdot,t))\right\rangle _{\mathbf{S}}+\eta\bigtriangleup o(\mathbf{x},t)\,,
\label{eq:dynamics_od}
\end{equation}
where
\begin{equation*}
A_{\sigma}(\mathbf{x},\mathbf{S},o(\cdot,t))=\frac{e^{-(|\mathbf{s}_{r}-\mathbf{x}|^{2}+|s_{o}-o(\mathbf{x},t)|^{2})/2\sigma^{2}}}{\int d^{2}y\, e^{-(|\mathbf{s}_{r}-\mathbf{y}|^{2}+|s_{o}-o(\mathbf{y},t)|^{2})/2\sigma^{2}}}
\end{equation*}
is the cortical activity pattern, $\sigma$ controls the receptive
field size in the stimulus parameter space, $\left<\cdot\right>$
denotes the average of the ensemble of visual stimuli \{${\mathbf{S}}$\},
$\eta$ measures the strength of lateral interactions and $\Delta$
is the two-dimensional Laplacian.  Visual stimuli $\mathbf{S}=(\mathbf{s}_{r},s_{o})$
are point-like and characterized by a location $\mathbf{s}_{r}$ and
an OD value $s_{0}$, which describes whether the activated units
are forced to prefer the ipsilateral ($s_{o}<0$) or the contralateral
($s_{o}>0$) eye.
\newline
\textbf{Numerical integration.}
Simulations were performed on a 64\texttimes{}64 grid with periodic
boundary conditions. We used at least 4 grid points per $\Lambda_{\textnormal{max}}$ and 
an integration time step $\delta t=\min\left\{ 1/(20\eta k_{\max}^{2}),\tau/10\right\} $.
The first term on the r.h.s. of Eq. \ref{eq:dynamics_od} was treated by an Adams-Bashforth scheme, 
the second term by spectral integration.  
$\mathbf{s}_{r}$ and $s_{o}$ were uniformly distributed with $\left<s_{o}^{2}\right>=1$.
Typical, between  4\texttimes10$^4$ and 2\texttimes10$^5$  stimuli were used per integration step. 
\newline
\textbf{Instantaneous Area Increases.} 
We rescaled the system length $L$  as determined from the 
desired value of $\delta k/k_{\textnormal{max}}$ (no change in number of grid points, see \textit{SI Appendix}).
We adjusted the number of stimuli, $N_{s}$, and, since $\Delta\sim1/L^{2}$,
the matrix for the spectral integration step. The numerical
value of $\sigma$ remained constant.
\newline
\textbf{Continuous Area Increases.} 
We linearly increased the linear extent $L$ of the simulated
regions between $t=0\tau$ and $t=100\tau$ and updated the Laplacian $\Delta$
and the number of stimuli $N_{s}$ at every integration
step.  

\subsection{Data Analysis Method}
Column spacing $\Lambda$ and bandedness $\alpha$ of both, data and
simulations, were analyzed using the wavelet-method introduced in
\cite{Kaschube:2002p429}. An over-complete basis of complex Morlet-wavelets at 
various scales and orientations was compared to the OD pattern at each spatial location. 
$\Lambda$ was estimated by the scale of the best matching wavelet, 
$\alpha$ by the angular variance of matching at that scale 
(see \textit{SI Appendix}).
\newline
\textbf{Statistics.}
r-values denote Pearson's linear correlation coefficient; p-Values were obtained with Student's t-tests.
\newline 
All methods are described in detail in the \textit{SI Appendix}.

\section*{Acknowledgements}

We express our gratitude to F. Wolf for inspiring discussions. We
thank M. Huang and F. Wolf for comments on numerical procedures. We
thank P. Mehta and S. Palmer for helpful comments on the manuscript.

\clearpage
\newpage


\section{SI Appendix}

\section{Optical imaging data}

\textbf{Surgery}. Anaesthesia was induced with an intramuscular injection
of ketamine (10 mg/kg Ketanest\textregistered{}, Parke-Davis, Berlin,
Germany) and xylazine hydrochloride (Rompun\textregistered{}, Bayer
AG, Leverkusen, Germany) and maintained throughout the experiment
using nitrous oxide/oxygen anaesthesia (50\% N$_{2}$O / 50\% O$_{2}$),
supplemented with halothane (0.8-1.2\%, Eurim Pharma, Germany). The
ECG, pulmonary pressure, end tidal CO$_{2}$ (3-4\%), and rectal temperature
(37-38\textdegree{}) were continuously monitored. The animal's head
was fixed in a stereotactic frame by means of a metal nut cemented
to the skull. For optical imaging of V1 a craniotomy was performed
centered at Horsley-Clarke coordinate P4. All experiments were performed
when the animals were between 28 and 94 days old. Successive experiments
in the same animal were performed with intervals of 7 days and in
one case 42 days. All animal experiments have been performed according
to the German Law on the Protection of Animals and the corresponding
European Communities Council Directive of November 24, 1986 (86/609/EEC).

\textbf{Visual Stimulation}. Animals were stimulated monocularly with
high-contrast square-wave gratings (subtending 90\textdegree{}\texttimes{}60\textdegree{}
visual field) of four orientations (0\textdegree{}, 45\textdegree{},
90\textdegree{}, and 135\textdegree{}) moving at a speed of 2cyc/s
with a spatial frequency of 0.5cyc/deg. Stimuli were generated by
EZV-Stim software (Optical Imaging Inc., Rehovot, Israel) and presented
on a LG Electronics Flatron 295 LCD-monitor (luminosity 180 cd/m\texttwosuperior{};
contrast 300:1; refresh rate 85 Hz; resolution 1600\texttimes{}1200
pixel) at a distance of 25 cm. The eyes were treated with atropine
and Neosynephrine\textregistered{} and refracted appropriately using
corrective corneal contact lenses with artificial pupils with a diameter
of 3 mm.

\textbf{Data Acquisition}. The cortical surface was illuminated by
means of two adjustable light guides attached to a tungsten-halogen
lamp (Spindler \& Hoyer, G\"ottingen, Germany) equipped with interference
filters for different wavelengths. The vascular pattern of the cortex
was visualized at 546 nm \textpm{} 10 nm (green), cortical activity
maps at 707 nm \textpm{} 1 nm (red). During data acquisition of intrinsic
signals, the camera was focused 650\textendash{}750 $\upmu$m below
the cortical surface. A tandem-lense was used for imaging \cite{ratzlaff_91}.
The ORA 2001 system (Optical Imaging Inc.), equipped with a cooled
Theta CCD system (384\texttimes{}288 pixel chip from Thomson-CSF)
was used for collecting the intrinsic signals. We acquired a series
of frames every 12 s, whereby a grating of a given orientation was
presented for 2 s in a static mode, followed by 4.2 s of data acquisition
during which the grating was moved in both directions along the axis
orthogonal to its orientation. We used episodic stimulation during
data acquisition (7 frames of 600 ms duration). The first frames were
excluded from further analysis. The stimulus presentation was monocular,
an eye shutter was used to conceal the eyes. A single stimulus trial
consisted of 2\texttimes{}8 stimulus conditions (4 grating orientations
for the left and the right eye) and 8 isoluminant blanks presented
in a random sequence. Twelve trials were usually presented to obtain
a map, so that every stimulus was shown 24 times. We first calculated
`single condition maps' in which the images acquired during
presentation of a particular stimulus were divided by the sum of all
different stimulus conditions (`cocktail blank procedure') \cite{bonhoeffer_96,engelmann_02}.
Differential maps for ocular dominance (OD) were calculated by summing
all left eye activity maps and subtracting all right eye activity
maps.

\textbf{Data Preprocessing}. All computed differential maps $I'(\mathbf{x})$ were preprocessed
in order to remove overall variations in signal strength and measurement
noise. We calculated a high-pass filtered map $I(\mathbf{x})=I'(\mathbf{x}) - J(\mathbf{x})$ by subtracting the regional mean
$$
J(\mathbf{x}) = \frac{1}{W(\mathbf{x})}\mathcal{F}^{-1}\left\{\tilde{K}_{hp}(\mathbf{k})\tilde{I'}(\mathbf{k})\right\}\,,
$$
where $\mathcal{F}$ denotes the Fourier transform and the Fermi-function 
$$
\tilde{K}_{hp}(\mathbf{k}) = \frac{1}{1+e^{-(k_{hp}-|\mathbf{k}|)/\beta_{hp}}}
$$
is parametrized by the high-pass cutoff frequency $k_{hp}$ and the steepness $\beta_{hp}$. After back-transformation to real space, the signal outside the region of interest (ROI) was discarded.  Normalizing by $W(\mathbf{x}) = \int _{ROI} d^2 x'\,K_{hp}(\mathbf{x} - \mathbf{x}')$ accounted for the boundary of the ROI. We used a Fermi filter with $\beta = 0.2k_{hp}$ and $k_{hp}=2\pi/\lambda_{hp}$ with $\lambda_{hp}=1.8$mm. Lowpass filtering was done with a second Fermi filter $\tilde{K}_{lp}(\mathbf{k})$ with parameters  $\beta = 0.2k_{lp}$ and $k_{lp}=2\pi/\lambda_{lp}$ with $\lambda_{lp}=0.7$mm.

The resulting pattern $I(\mathbf{x})$ was then centered to yield $\int_{ROI}d^{2}x\, I(\mathbf{x})=0$ and its variance was normalized to one. This overall bandpass filtering ensured that
structures on a scale between 0.7 mm and 1.8 mm were only weakly attenuated
by the preprocessing and enabled us to do further quantitative analysis. 
\section{Data analysis methods}
\noindent
\textbf{Column spacing.}
For each preprocessed OD pattern $I(\mathbf{x})$, we calculated a wavelet representation,
using wavelets which covered only a few hypercolumns but exhibited
a strong periodicity. These representations were obtained from\[
\hat{I}(\mathbf{x,}\theta,l)=\int_{ROI}d^2y\, I(\mathbf{y})\psi_{\mathbf{x},\theta,l}(\mathbf{y})\,,\]
where $\mathbf{x},\theta,l$ are the position, orientation and scale
of the wavelet $\psi_{\mathbf{x},\theta,l}$ and $\hat{I}(\mathbf{x,}\theta,l)$
denotes the array of wavelet coefficients. We used complex Morlet
wavelets defined by\[
\psi(\mathbf{x})=\exp\left(-\frac{(x_{1}^{2}+\sigma_2^{-2}x_{2}^{2})}{2}\right)e^{i\mathbf{k}_{\psi}\mathbf{x}}\]
and \[
\psi_{\mathbf{x},\theta,l}(\mathbf{y})=l^{-1}\psi\left(\Omega^{-1}(\theta)\frac{\mathbf{y}-\mathbf{x}}{l}\right)\,,\]
where \[
\Omega(\theta)=\left(\begin{array}{cc}
\cos\theta & -\sin\theta\\
\sin\theta & \cos\theta\end{array}\right)\]
is the two-dimensional rotation matrix.
The characteristic wavelength of a wavelet with scale $l$ is $\Lambda_{\psi}l$
with $\Lambda_{\psi}=2\pi/|\mathbf{k}_{\psi}|$. $\sigma_2$ denotes
the anisotropy of the wavelet. We used relatively large 
isotropic wavelets ($\mathbf{k}_{\psi}=(7,0)$, $\sigma_2=1$) to estimate local column spacing $\Lambda(\mathbf{x})$.
We used 12 equally spaced orientations and 16 scales $l_{j}$ between
0.5 mm and 2.0 mm. For computational efficiency, 9 equally spaced orientation were used to analyze model cortices. Test analyses with 12 orientation led to almost indistinguishable results. 
We first calculated \[
\bar{I}(\mathbf{x},l)=\int_{0}^{\pi}\frac{d\theta}{\pi}|\hat{I}(\mathbf{x},\theta,l)|\]
 of the wavelet coefficients for every position $\mathbf{x}$ and
then determined the scale by computing $\bar{l}(\mathbf{x})=\operatorname{arg\,max}_{\,l}(\bar{I}(\mathbf{x},l))$.
The corresponding characteristic local wavelength was obtained by $\Lambda(\mathbf{x})=l(\mathbf{x})\Lambda_{\psi}$.
  $\bar{l}(\mathbf{x})$ was estimated as the
maximum of a polynomial of 6th degree in $l$, fitting the $\bar{I}(\mathbf{x},l_{j})$
(least square fit) for a given position $\mathbf{x}$. 
Based on the local column spacing $\Lambda(\mathbf{x})$, we calculated the mean
column spacing $\Lambda=\left<\Lambda(\mathbf{x})\right>_{x}$, where here and in the following, $\left<\cdot\right>_{x}$ denotes the average over the ROI (e.g. V1 or simulated domain). In the case of 2-DG maps, we analyzed column spacings in $N=4$ different brain slices for each hemisphere and averaged over slices to obtain more accurate estimations of the mean column spacing \cite{Kaschube:2002p429,Kaschube:2003p430}. This procedure also provides an estimation of the measurement error. On average errors  were below 0.02mm implying a  relative error (error/column spacing) of less than 2\%. For the optical imaging data, errors were estimated by bootstrap resampling. For each map, a bootstrap sample of N=100 maps was generated from the individual trials. The column spacing was calculated for each map in the sample. The error was estimated by the standard deviation of the bootstrap distribution of column spacings.

\noindent
\textbf{Number of hypercolumns.}
We defined the number of hypercolumns $N_{HC}$ in an area $A$ by $N_{HC}=A/\Lambda^{2}$, where $\Lambda=\left<\Lambda(\mathbf{x})\right>_{x}$ is the mean column spacing in A.
 Note that we do not assume a specific shape of the hypercolumn, but solely that its size is equal to $c \Lambda^2$ where $c$ is a constant close to 1. For simplicity, and because we are primarily interested in relative changes in $N_{HC}$, we set $c=1$.  
Note further that for simplicity we are assuming  $\left<\Lambda(\mathbf{x})\right>^{-2}_{x}\approx \left<\Lambda(\mathbf{x})^{-2}\right>_{x}$. 
Typical standard deviations  $\sigma_{\Lambda}=\sqrt{\left<\Lambda(\mathbf{x})^2\right>_{x}}$ of OD column spacings in the cat are $\approx0.1\Lambda$ \cite{Kaschube:2009} and in the model $<0.03\Lambda$. Based on these values, this approximation is accurate up to  $\approx 3\%$.

\noindent
\textbf{Bandedness.}
\begin{figure}[htbp]
\begin{center}
\includegraphics[width=9cm]{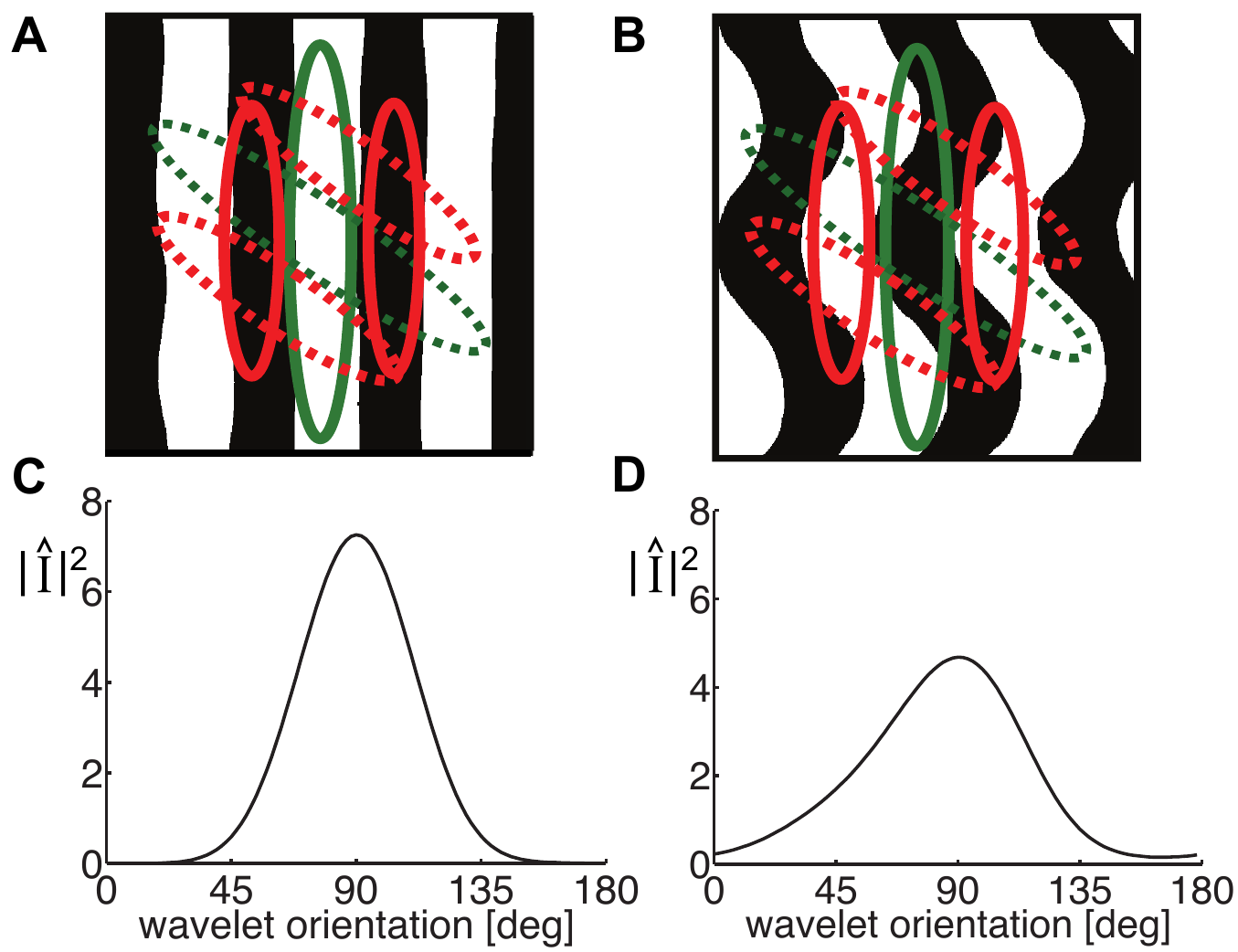}
\end{center}
\caption{{\bf Wavelet analysis of columnar layouts.} (A) Examples of two
wavelets superimposed on a stripe-like region of a simulated OD pattern.
The real parts of the complex-valued wavelet are depicted. Positive
regions are delineated by red lines, negative regions are delineated
by green lines. The two  wavelets exhibit the best matching spatial frequency. One wavelet 
(solid line) also exhibits an optimal orientation.
 (B) Wavelets superimposed on sinusoidally bended stripe
patterns, characteristic for the ZZ reorganization. (C-D)
The normalized squared modulus of the wavelet coefficients as a function
of orientation $\theta$ for the pattern in A (C)
and for the pattern in B (D). Note that for the
stripe-like pattern $|\hat{I}(\mathbf{x},\theta)|^{2}$ is strongly
modulated and exhibits a pronounced peak, whereas the sinusoidally bended
stripes lead to a broader $\theta$-dependence and a less pronounced
peak. This difference is captured by the bandedness parameter $\alpha$. \label{fig:wavelet_sheme} }
\end{figure}
The orientation dependence of the wavelet coefficients was used to calculate a parameter
measuring the anisotropy of local pattern elements as shown in Fig.
S\ref{fig:wavelet_sheme}. For a pattern consisting of parallel stripes the magnitude of the wavelet coefficients depends strongly
on the wavelet orientation and is largest if the orientation of the
wavelet matches the orientation of the bands (Fig. \ref{fig:wavelet_sheme}, A and
C). For a pattern consisting of more bended stripes or isotropic
patches, the wavelet coefficients depend only weakly on the orientation
of the wavelet (Fig. \ref{fig:wavelet_sheme}, B and D).  Therefore, using
only wavelets of wavelength $\Lambda(\mathbf{x})$, we 
calculated
\begin{equation}
s'(\mathbf{x})=\int_{0}^{\pi}d\theta|\hat{I}(\mathbf{x},\theta)|^{2}e^{i2\theta}\left/\int_{0}^{\pi}d\theta|\hat{I}(\mathbf{x},\theta)|^{2}\right.\,,\label{eq:bandedness_1}\end{equation}
where we used relatively small and anisotropic wavelets ($\mathbf{k}_{\psi}=(2,0)$, $\sigma_2=1.5$) in order to resolve the layout locally.  
We defined the local bandedness as\begin{equation}
s(\mathbf{x})=\int_{ROI}d^{2}y\, K(\mathbf{x}-\mathbf{y})s'(\mathbf{y})\left/\int_{ROI}d^{2}y\, K(\mathbf{x}-\mathbf{y})\right.\,,\label{eq:bandedness_2}\end{equation}
where $K(\mathbf{x})=\frac{1}{2\pi\sigma^{2}}\exp\left(-\mathbf{\mathbf{x}^{2}}/2\sigma^{2}\right)$
and $\sigma=1.3\Lambda$. To estimate \eqref{eq:bandedness_1} and
\eqref{eq:bandedness_2}, wavelet coefficients were computed for 9
equally spaced orientations for both model cortices and data.  Test analyses with 12 equally spaced orientations led to almost indistinguishable results.
With the above choice of $\sigma$, the function $s(\mathbf{x})$
is sensitive to the occurrence of band-like regions that extend over
the size of more than a hypercolumn. Based on the local bandedness $s(\mathbf{x})$, 
we computed the bandedness $\alpha=\left<s(\mathbf{x})\right>_{x}$
to characterize the overall layout of simulated and measured OD patterns.
To estimate measurement errors for 2-DG maps, the bandedness was calculated for $N=4$ brain slices and then averaged \cite{Kaschube:2002p429,Kaschube:2003p430}. Relative errors for bandedness estimation were below $\approx 5\%$.
\section{The Elastic Network model - linear stability analysis and numerical procedures}
\subsection{Model definition and linear stability analysis}
In the continuous version \citep{Wolf:1998p1199} of the Elastic Network (EN) model \citep{Durbin:1990p1196,Goodhill:2000p2141},
the OD map in V1 at a given time $t$ is described by a real valued
field $o(\mathbf{x},t)$, where $\mathbf{x}$ represents the position
on the cortical surface. Negative values of $o(\mathbf{x},t)$ indicate
a preference for input from the ipsilateral eye, positive values a
preference for the contralateral eye. This field follows a dynamics
\begin{equation}
\partial_{t}o(\mathbf{x},t)=\left\langle \left[s_{o}-o(\mathbf{x},t)\right]A_{\sigma}(\mathbf{x}, \mathbf{S},o(\cdot,t))\right\rangle _{\mathbf{S}}+\eta\bigtriangleup o(\mathbf{x},t)\,,
\label{eq:dynamics_od_suppl}
\end{equation}
where
\begin{equation}
A_{\sigma}(\mathbf{x},\mathbf{S},o(\cdot,t))=\frac{e^{-(|\mathbf{s}_{r}-\mathbf{x}|^{2}+|s_{o}-o(\mathbf{x},t)|^{2})/2\sigma^{2}}}{\int d^{2}y\, e^{-(|\mathbf{s}_{r}-\mathbf{y}|^{2}+|s_{o}-o(\mathbf{y},t)|^{2})/2\sigma^{2}}}
\label{eq:activation_function_OD_Suppl}
\end{equation}
is the cortical activity pattern, $\sigma$ controls the receptive
field size in the stimulus parameter space, $\left<\cdot\right>_{\mathbf{S}}$
denotes the average of the ensemble of visual stimuli \{${\mathbf{S}}$\},
$\eta$ measures the strength of lateral interactions and $\Delta$
is the two-dimensional Laplacian. Selectivities $o(\mathbf{x})$ are
modified through the cumulative effect of a large number of activity
events, evoked by the complete stimulus ensemble. Visual stimuli $\mathbf{S}=(\mathbf{s}_{r},s_{o})$
are point-like and characterized by a location $\mathbf{s}_{r}$ and
an OD value $s_{0}$ which describes whether the activated units
are forced to prefer the ipsilateral ($s_{o}<0$) or the contralateral
($s_{o}>0$) eye. 
The stimulus parameters $\mathbf{s}_r$ and $s_{o}$ are uniformly distributed
with densities $\rho_{\mathbf{s}_r}$ and $\rho_{s_{o}}$ such that $\left<s_{o}^{2}\right>=1$.
\newline

\noindent
\textbf{Linear stability analysis around nonselective fixed point.} 
By linear stability analysis, we show in the following that OD columns segregate by
a finite wavelength instability. We linearize eq. (\ref{eq:dynamics_od_suppl})
around the homogeneous nonselective state $o(\mathbf{x})=0$  \citep{Wolf:1998p1199}.
Inserting $o_{\hom}(\mathbf{x})\equiv0$ into eq. (4)
yields
\begin{eqnarray*}
A_{\sigma}(\mathbf{x},\mathbf{\, S},\, o_{\hom}) & = & \frac{e^{-\frac{|\mathbf{s}_{r}-\mathbf{x}|^{2}}{2\sigma^{2}}}}{2\pi\sigma^{2}}\,.
\end{eqnarray*}
After averaging over the ensemble of stimuli with normalized uniform
densities $\rho_{\mathbf{s}_r}$ and $\rho_{s_{o}}$, we obtain\[
\left.\partial_{t}o(\mathbf{x})\right|_{o_{\hom}}=\int d^2s_{r}\,\rho_{\mathbf{s}_r}\frac{e^{-\frac{|\mathbf{s}_r-\mathbf{x}|^{2}}{2\sigma^{2}}}}{2\pi\sigma^{2}}\int ds_{o}\,\rho_{s_{o}}s_{o}=0\,.\]
This shows that $o_{\hom}(\mathbf{x})\equiv0$ is a fixed point of
the dynamics in eq. (\ref{eq:dynamics_od_suppl}). The stability of this
fixed point can be studied by linearizing the r.h.s of eq. (\ref{eq:dynamics_od_suppl})
\citep{Wolf:1998p1199,Scherf:1999p839}, \[
\mathcal{F}[o,\mathbf{x},\sigma,\mathbf{S}]\approx\mathcal{F}[o_{\hom},\mathbf{x},\sigma,\mathbf{S}]+\int d^{2}y\left.\frac{\delta\mathcal{F}}{\delta o(\mathbf{y})}\right|_{o_{\hom}}o(\mathbf{y})\,.\]
This yields\begin{eqnarray*}
\partial_{t}o(\mathbf{x}) & \approx & \eta\Delta o(\mathbf{x})-o(\mathbf{x})\iint d^{2} s_r\, ds_{o}\, d^{2}y\,\rho_{s_{o}}\rho_{\mathbf{s}_r}A_{\sigma}(\mathbf{y},\mathbf{S},o_{\hom})\\
 &  & +o(\mathbf{x})\iint d^{2}s_r\, ds_{o}\, d^{2}y\,\rho_{s_{o}}\rho_{\mathbf{s}_r}A_{\sigma}(\mathbf{y},\mathbf{S},o_{\hom})\frac{s_{o}^{2}}{\sigma^{2}}\\
 &  & -\iint d^{2} s_r\, ds_{o}\, d^{2}y\,\rho_{s_{o}}\rho_{\mathbf{s}_r}\frac{s_{o}^{2}}{4\pi^{2}\sigma^{6}}e^{-\left(|\mathbf{s}_r-\mathbf{x}|^{2}+|\mathbf{s}_r-\mathbf{y}|^{2}\right)/2\sigma^{2}}o(\mathbf{y})\,.\end{eqnarray*}
Assuming a uniform stimulus density $\rho_{\mathbf{s}_r}$ across the cortical
surface, this results in\begin{equation}
\partial_{t}o(\mathbf{x})=\left(\eta\bigtriangleup+\frac{\left<s_{o}^{2}\right>}{\sigma^{2}}-1\right)o(\mathbf{x})-\frac{\left<s_{o}^{2}\right>}{4\pi\sigma^{4}}\int d^{2}y\, e^{-\frac{\left(\mathbf{x}-\mathbf{y}\right)^{2}}{4\sigma^{2}}}o(\mathbf{y})\,.\label{eq:linearization}\end{equation}
Note that the linearization is only governed by the variance
of the stimulus ensemble, and not by higher order statistical moments.

\begin{figure*}[htbp]
\begin{center}
\includegraphics[width=16cm]{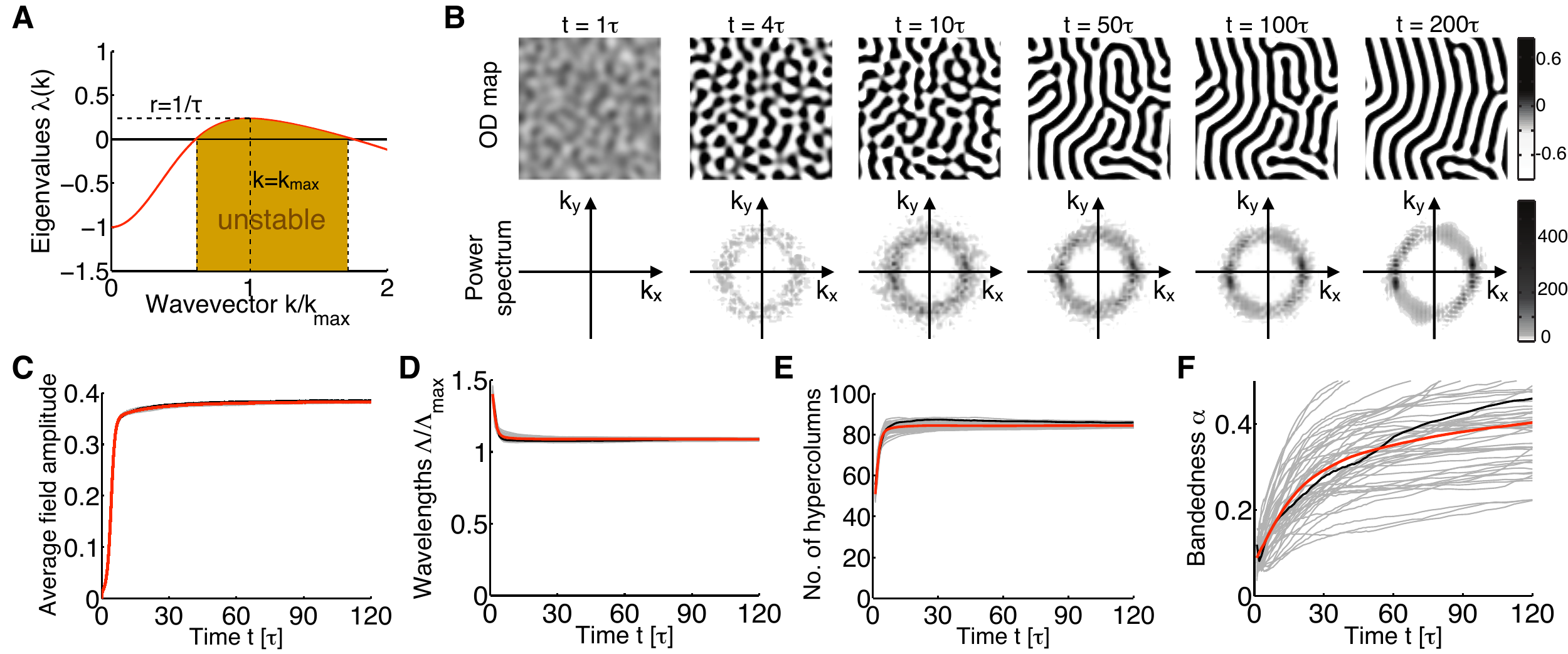}
\end{center}
\caption{{\bf  Development of OD column layout in the Elastic Network model without
growth.} ({A}) Eigenvalues of the linearized dynamics around
the homogeneous nonselective state, $o(\mathbf{x})=0$. For maximal eigenvalue $r(\sigma,\eta)>0$,
Fourier modes with positive growth rate form a band around $k_{\max}$, 
corresponding to an annulus in 2D. OD columns with spatial scale
$\Lambda_{\max}=2\pi/k_{\max}$ segregate on a time scale
$\tau=1/r$. ({B}) Development
of OD columns starting from initial condition $o(\mathbf{x},t=0)=0$
($\eta=0.025$, $r=0.2$). At $t=4\tau$, the emerging pattern already
exhibits characteristics of OD columns. After $10\tau$, columns start
to merge and progressively reorganize towards a stripe-like layout.
({C-F}) Time courses of average OD segregation $\mathcal{A}$
({C}) and of three parameters, characterizing OD layouts during
development: ({D}) mean column
spacing $\Lambda$ ({E}) number of hypercolumns $N_{HC}$ ({F})
mean bandedness $\alpha$ (N = 50 realizations, gray: individual traces, black:
example from {B}, red: mean value). Whereas $\mathcal{A}$, $\Lambda$, and
$N_{HC}$ reach mature levels around 10$\tau$ and exhibit only little
variability, the bandedness $\alpha$ increases during the entire
time course, expressing the fact that\textit{\emph{ solutions slowly
converge towards}} ideal OD \textit{\emph{stripes}}. Furthermore,
$\alpha$-time courses display a large variability, capturing differences
between individual realizations of OD map development. \label{fig:elastic_net_model}}
\end{figure*}

Due to translation and rotation invariance of the
model, the eigenfunctions of its linearized dynamics are Fourier
modes $\sim e^{i\mathbf{kx}}$ with eigenvalues \cite{Wolf:1998p1199,Scherf:1999p839}
\begin{equation*}
\lambda(k)=-1+\frac{\left<s_o^2\right>}{\sigma^2}\left(1-e^{-k^2\sigma^2}\right)-\eta k^2\label{eq:spectrum_EN}
\end{equation*}
only depending on the absolute value $|\mathbf{k}|=k$ 
(see Fig. 2A for an illustration). 
The homogeneous nonselective state $o_{\hom}(\mathbf{x})\equiv0$ is
unstable, if some eigenvalue with $k>0$ is larger than zero. 
The amplitude of any small perturbation
containing a Fourier mode with wave vector $\mathbf{k}$ will evolve
$\sim\exp(\lambda(k)t)$, and therefore spatial frequencies with $\lambda(k)<0$
are exponentially damped, whereas those with $\lambda(k)>0$ grow
exponentially (shaded region in Fig. \ref{fig:elastic_net_model}A).
For $\eta>0$ and $\sigma>0$, $\lambda(k)$ has a single 
maximum at $k_{\max}=\frac{1}{\sigma}\sqrt{\ln(1/\eta)}$. The maximum
$\lambda(k_{\max})$ is positive if the width $\sigma$
of the activation function is smaller than $\sigma^{*}=\sqrt{1-\eta+\eta\ln\eta}\,.$
Importantly, the maximum positive eigenvalue $r=\lambda(k_{\max})$
defines a time scale $\tau=1/r$ on which OD columns segregate.
We refer to $r$ as the control parameter. 
For $r>0$, the homogeneous nonselective state is unstable and the maximum is at some finite wavelength 
$\Lambda_{\max}=2\pi/k_{\max} $.
This spatial scale is roughly the spacing of columns in the developing OD map. 
Confirming this linear stability analysis, we find numerically that the early pattern consists of Fourier modes with wavelength $\approx\Lambda_{\max}$ (see below). 
Thus, in the EN model, OD columns segregate because the
nonselective state $o(\mathbf{x})=0$ becomes unstable and Fourier
modes with period $\sim\Lambda_{\max}$ grow exponentially
if $\sigma$ is below a critical value. 
\newline

\noindent
\textbf{Generality of model definition.}
In the above definition of the Elastic Network model, the widths of the activation function $A_{\sigma }(\mathbf{x},\mathbf{S},o(\cdot,t))$ in OD space and retinotopic space are identified and set to $\sigma$. In the following, we show that this can always be achieved by a proper rescaling of cortical space, and hence does not  imply a loss of generality.

With two different widths, $\sigma_r$ for retinotopic space and $\sigma_o$ for OD, the EN model becomes
\begin{equation}
\partial_{t}o(\mathbf{x},t)=\left\langle \left[s_{o}-o(\mathbf{x},t)\right]A_{\sigma_o,\sigma_r }(\mathbf{x}, \mathbf{S},o(\cdot,t))\right\rangle _{\mathbf{S}}+\eta\bigtriangleup o(\mathbf{x},t)\,,\label{eq:dynamics_od_general_suppl}
\end{equation}
where\begin{equation*}
A_{\sigma_o,\sigma_r }(\mathbf{x},\mathbf{S},o(\cdot,t))=\frac{e^{-((|\mathbf{s}_{r}-\mathbf{x}|^{2})/2\sigma_r^2+(|s_{o}-o(\mathbf{x},t)|^{2})/2\sigma_o^{2})}}{\int d^{2}y\, e^{-((|\mathbf{s}_{r}-\mathbf{x}|^{2})/2\sigma_r^2+(|s_{o}-o(\mathbf{x},t)|^{2})/2\sigma_o^{2})}}\,.
\label{eq:activation_function_OD_general}
\end{equation*}
For the spectrum of eigenvalues of the linearized dynamics we obtain
\begin{equation*}
\lambda(k)=-1 + \frac{\left<s_o^2\right>}{\sigma_o^2}\left(1-e^{-k^2\sigma_r^2}\right) - \eta k^2\,.
\end{equation*}
and hence the critical wavenumber (i.e. the expected wavenumber of the emerging OD pattern) is given by
\begin{equation*}
k_c = \frac{1}{\sigma_r}\sqrt{\ln\left(\frac{\left<s_o^2\right>}{\sigma_o^2}\frac {\sigma_r^2}{\eta}\right)}\,.
\end{equation*}
By a rescaling of cortical space according to
\begin{eqnarray*}
\mathbf{x} & \rightarrow &  \mathbf{x}'= \alpha\mathbf{x}\\
\mathbf{k} & \rightarrow &  \mathbf{k}'= \frac{1}{\alpha}\mathbf{k}\\
\sigma_r & \rightarrow &  \sigma_r' = \alpha \sigma_r \\
\eta & \rightarrow &\eta' =   \alpha^2\eta\, ,
\end{eqnarray*}
where the last transformation rescales the laplacian term in the r.h.s of eq. (\ref{eq:dynamics_od_general_suppl}), we obtain
$$
k_c'  = \frac{1}{\alpha }k_c
$$
and hence $\Lambda_{\max}' = \alpha \Lambda_{\max}$. The typical column spacing of the emerging OD pattern is properly transformed under the above rescaling. Importantly, the maximal growth rate
\begin{eqnarray*}
r' = \frac{1}{\tau'} = \lambda'(k_c') &= &-1 + \frac{\left<s_o^2\right>}{\sigma_o^2}\left(1-e^{-k_c'^2\sigma_r'^2}\right) - \eta' k_c'^2 \\
					 & = &\lambda(k_c) = \frac{1}{\tau} = r 
\end{eqnarray*}
is unaffected by the scaling transformation. In fact, one can show that not only the linearized dynamics, but also all higher order terms determining pattern selection, are unchanged by this rescaling. 

These considerations imply that the numerical value of $\sigma_r$ can always be identified with the value of $\sigma_o$ without loss of generality. In a final step, we choose to measure space in dimensionless units which allows for setting $\sigma_o = \sigma_r = \sigma$ and we arrive at the model definition as given in eq. (1) of our manuscript. 

The linear as well as higher order terms exclusively depend on the ratio between $\left<s_o^2\right>$ and $\sigma_o^2$, meaning that $\left<s_o^2\right> $ just sets an arbitrary scale in OD space. For convenience, we chose $\left<s_o^2\right>=1$ in our analytics as well as in our numerics. 

In our simulations, we have applied the opposite direction of the above argumentation: fixing $r$ and $\eta$ (e.g., when computing the diagram in Fig. 3C), implied a certain ratio between $\left<s_o^2\right>/\sigma^2$ as well as a numerical value of $\Lambda_{\max}$. The numerical value, $L$, of the system size was then set to a multiple of  $ \Lambda_{\max}$,  $L = \gamma \Lambda_{\max}$, with $\gamma\in \mathbb{R}$ chosen according to the size of the cortical subregion (in terms of hypercolumns) we 
wanted to simulate (see below).

\subsection{Numerical procedures}
Simulations were performed on a 64\texttimes{}64 grid with periodic
boundary conditions. Simulated systems were spatially discretized
with at least 4 grid points per $\Lambda_{\max}$ to achieve
sufficient resolution in space. Test simulations with larger grid
sizes (128\texttimes{}128, 256\texttimes{}256) did not lead to significantly
different results. Progression of time was measured in units of the
intrinsic time scale $\tau$. Here, the integration time step $\delta t$
is bounded by the relevant decay time constant of the Laplacian in
eq. \eqref{eq:dynamics_od_general_suppl} around $k_{\max}$ and by the intrinsic
time scale $\tau$ of the system. We used $\delta t=\min\left\{ 1/(20\eta k_{\max}^{2}),\tau/10\right\} $
to ensure good approximation to the temporally continuous changes
of the OD patterns. Note that, since $\tau=1/r$, simulation time
diverges for $r\rightarrow0$. The EN model dynamics was simulated
using an Adams-Bashforth scheme for the first term on the r.h.s. of
eq. \eqref{eq:dynamics_od_general_suppl}. The second term was treated by spectral
integration, exhibiting unconditional numerical stability.  
To approximate the stimulus ensemble, a large random sample of pointlike
stimuli was drawn at each time step.
Different realizations of OD development were obtained by presenting
different stimulus samples.
The stimulus parameters $\mathbf{s}_r$ and $s_{o}$ were chosen to be uniformly distributed
with densities $\rho_{\mathbf{s}_r}$ and $\rho_{s_{o}}$ such that $\left<s_{o}^{2}\right>=1$ (see model definition above).
The stimulus average in eq. (\ref{eq:dynamics_od_general_suppl}) was approximated
by choosing a random representative sample of $N_{s}$ stimuli at
each integration time step, with\[
N_{s}=\frac{N_{0}\Gamma^{2}}{\delta t}\varepsilon_{s}^{-n}\sqrt{\tau g_{00}}\,,\]
where $n$ is the number of dimensions of the feature space (in our case, $n=3$),
$\Gamma^{2}=(L/\Lambda_{\max})^{2}$ the size of the simulated
system in units of $\Lambda_{\max}^{2}$, $\varepsilon_{s}$
the resolution in feature space, $N_{0}$ the number of stimuli we
required to sufficiently approximate the cumulative effect of the
ensemble of stimuli within each feature space voxel, and $g_{00}=\frac{9(\eta-1)^{4}}{40\sigma^{6}}$
a factor that depends on the specific form of nonlinear competition
between Fourier modes in the EN model. $\sqrt{\tau g_{00}}$ is
proportional to the inverse of the expected mean amplitude of the
OD pattern. With $N_{0}=50$ and $\varepsilon_{s}=0.15$, 
we ensured a low amplitude to noise ratio for all the simulations.
Typical values for $N_{s}$ were between 40000 and 200000. To model
development prior to OD segregation, we initialized simulations with
$o(\mathbf{x},t=0)=0$.
\section{Formation of ocular dominance in the Elastic Network model without growth.} 
In this section, we briefly summarize the behavior of the Elastic Network model without cortical expansion. 
Fig. \ref{fig:elastic_net_model}B shows a typical simulation of OD development ($\eta = 0.025$, $r = 0.2$, no increase in area). A structure strongly resembling the OD pattern observed in cat V1 emerges after a few time steps. 
Already at $t=4\tau$ the OD pattern exhibits a visible characteristic spacing which becomes more
dominant in the later time course. 
As a measure of the average cortical selectivity, we monitored the mean field amplitude
$\mathcal{A}(t)=\int d^{2}x\,|o(\mathbf{x},t)|$. It shows a rapid increase during the first 10$\tau$ and saturates thereafter (Fig. \ref{fig:elastic_net_model}C). 

We quantified each simulated OD development by the wavelet-method
applied to our experimental data and compute three parameters characterizing
the map layouts. First, we estimated the mean column spacing $\Lambda$.
We find, that it reaches a constant value after approximately 10$\tau$
close to the predicted value $\Lambda_{\max}$ (Fig.
S\ref{fig:elastic_net_model}D). Next, we computed the number of hypercolumns
$N_{HC}$ in a simulated area. This number changes only very little
after OD columns have segregated (Fig. \ref{fig:elastic_net_model}E).
Finally, we calculated the bandedness, $\alpha$, which quantifies
the tendency of OD domains to form elongated parallel stripes. In all our simulations, the initial
phase of OD segregation is followed by a long stage of slow rearrangements
leading to progressively more stripe-like OD layouts. This behavior
is captured by a monotonically increasing bandedness $\alpha$ (Figure
S\ref{fig:elastic_net_model}F). In fact, we numerically find that
OD stripes, for which the bandedness is maximal, are stable
solutions of the EN model. We do not find any other stable solution. 
%
%
%
%
%
\section{Simulating growth}
\subsection{Instantaneous area increases}
We started simulations with
a predefined system length of $L_{0}$ and used a parameter-dependent
sine-wave initial condition
$$o(\mathbf{x})=1/\sqrt{\tau g_{00}}\sin(k_{\max}x)
$$
with periodicity $\Lambda_{\max}=2\pi/k_{\max}$
and amplitude close to the amplitude of the expected stationary state.
The system was integrated keeping the length of the system fixed for
the first 10$\tau$. Within this period the amplitude $\mathcal{A}(t)=\int d^{2}x\,|o(\mathbf{x},t)|$
of the OD stripe patterns relaxed towards a stationary value while
OD layout underwent only minor random fluctuations. Next, we rescaled
the system length to $L_{0}+\delta L$ (no change in number of grid
points), where
\[
\delta L=\frac{(\delta k/k_{\max})}{(\delta k/k_{\max})-1}L_{0}
\]
is determined from the desired value of $\delta k/k_{\max}$,
thereby increasing ($\delta L>0$) or decreasing ($\delta L<0$) the
total area size covered by the OD pattern. Accordingly, we adjusted
the number of stimuli, $N_{s}$, and, since $\Delta\sim1/L^{2}$,
the matrix for the spectral integration step. Note that the numerical
value of $\sigma$, i.e. the width of the activation function $A_{\sigma}(\mathbf{x},\mathbf{S},o(\cdot,t))$,
remains unchanged by this rescaling, but its extension relative to
$L$ decreases (and the number of grid points per activation blob
decreases as well). After rescaling, we continued simulations for
another 190$\tau$. 
\begin{figure*}[htbp]
\begin{center}
\includegraphics[width=14cm]{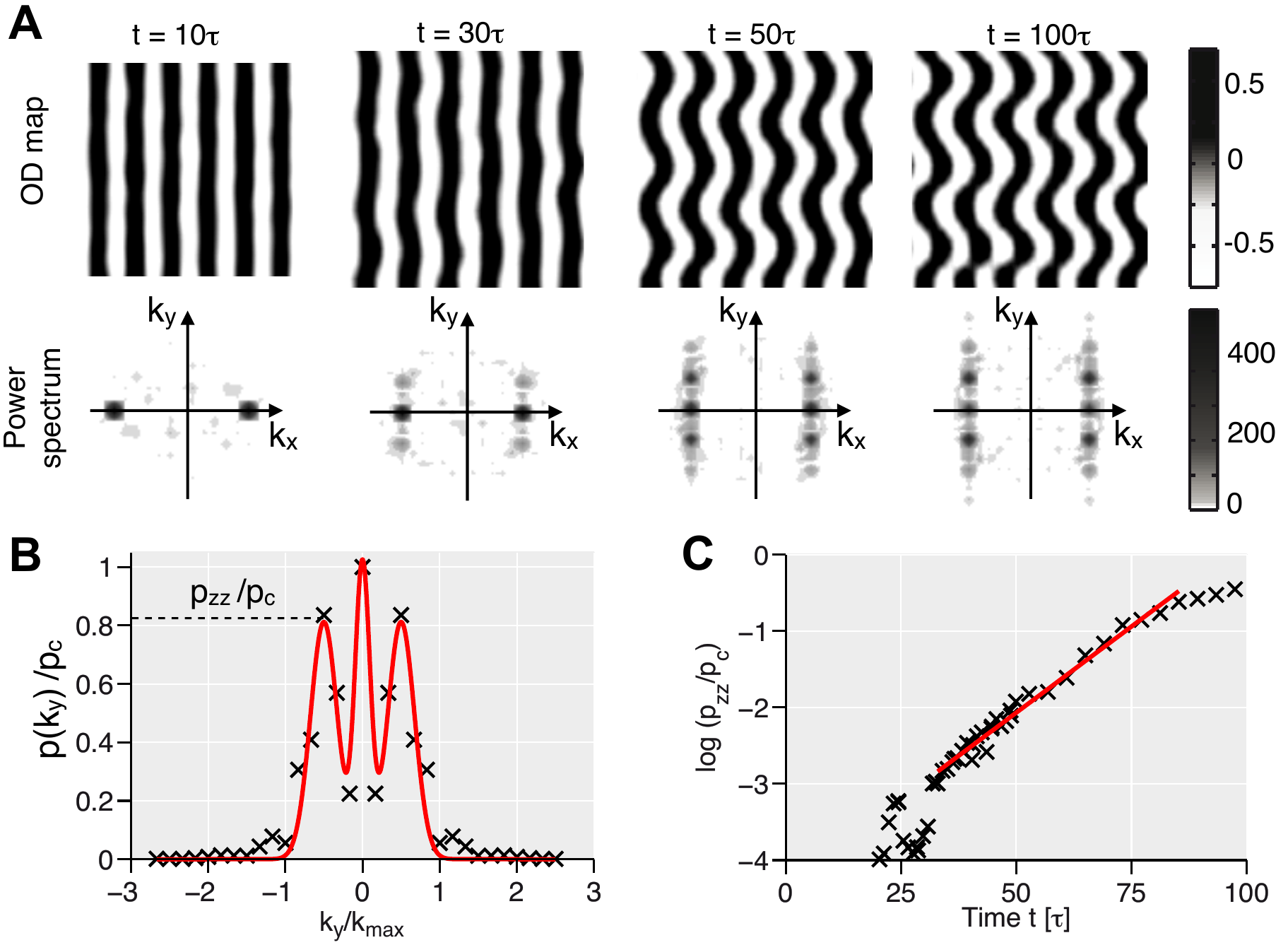}
\end{center}
\caption{ {\bf  Measuring the time scale of ZZ reorganization in the EN model.}
 (A) Snapshots of a simulation starting from a near
steady state solution of the Elastic Network (EN) model \cite{Durbin:1990p1196}, 
a stripe-like OD pattern (upper row) corresponding to a single mode in the 
power spectrum (lower row) ($\eta=0.025,\, r=0.15$) (redrawn from Fig. 2A in the manuscript).
After instantaneous area increase (by a factor of 1.18 (linear extent), i.e. $\delta k/k_{\max}=-0.15$, at 10$\tau$), OD domains bend sinusoidally and additional modes appear
at $(k_{\max}+\delta k)\mathbf{\vec{x}}\pm q_{y}\mathbf{\vec{y}}$.
(B) Measuring the growth rate of zigzag modes for the simulation in A. Black crosses
mark the power spectrum $p(k_{x} = k_{\max} + \delta k,\, k_{y})$.
Most power is concentrated around the origin $k_{y}=0$, stemming
from the initial stripe-like OD pattern, and around two secondary peaks corresponding
to the emerging ZZ modes (compare to A, lower row).
Red curve is a fit of three Gaussians (least square fit).
(C) Ratios between the height $p_{c}$ of the central
peak and height $p_{ZZ}$ of the side peaks at different time
points (black crosses). The growth rate of ZZ modes is estimated by fitting a linear slope to the
logarithm of $p_{ZZ}/p_{c}$ in the region of exponential growth (least square fit). \label{estimate_tau_zz}}
\end{figure*}
Upon instantaneous isotropic area increase, OD stripes typically display a zigzag-like bending of domains (Fig. \ref{estimate_tau_zz}A). 
As outlined below (sec. 6.3), a ZZ instability is characterized by the growth of two Fourier modes (representing the zig and the zag respectively) with wave vectors $\tilde{\mathbf{k}}_{ZZ} = (k_{\max}+\delta k)\tilde{\mathbf{x}}\pm q_{y}\tilde{\mathbf{y}}$, where $k_{\max}\mathbf{\vec{x}}$ is the wave vector of the original stripe pattern. 
A ZZ instability can therefore be reliably identified by monitoring the power spectrum along the axis $k_x = k_{\max}+\delta k$ and searching for two characteristic peaks corresponding to the ZZ modes.

To estimate the time scales $\tau_{ZZ}$ of this reorganization, we followed the growth of the two ZZ modes over time (see Fig. \ref{estimate_tau_zz}B and C). 
At each time point, we fitted Gaussians (least square fit) of variable size, width, and position to the original
mode and the two growing ZZ modes along the expected axis of the ZZ
modes in Fourier space (Fig. \ref{estimate_tau_zz}B).
The time scale of ZZ reorganization was extracted by linear fitting
(least square fit) the logarithm of the ratio between the peak height
of the ZZ modes, $p_{ZZ}$, and the central peak height stemming from
the original pattern, $p_{c}$ (Fig. \ref{estimate_tau_zz}C).
We constrained the fit to values between 15\% and 60\% of the maximal
peak ratio to capture best the regime of exponential growth of the
ZZ modes. The fitting procedure was only applied to simulations with
maximal peak ratio of at least 0.05. 
In Figure 3C of the manuscript, the inverse of the estimated growth rate, i.e. the time scale of the ZZ instability, $\tau_{ZZ}$, is compared to the intrinsic time scale $\tau$ of OD segregation for various changes of area ($\delta k$) and 
different values of the control parameter $r$.
\subsection{Continuous area increases}
\begin{figure*}[htbp]
 \centering \includegraphics[width=13cm]{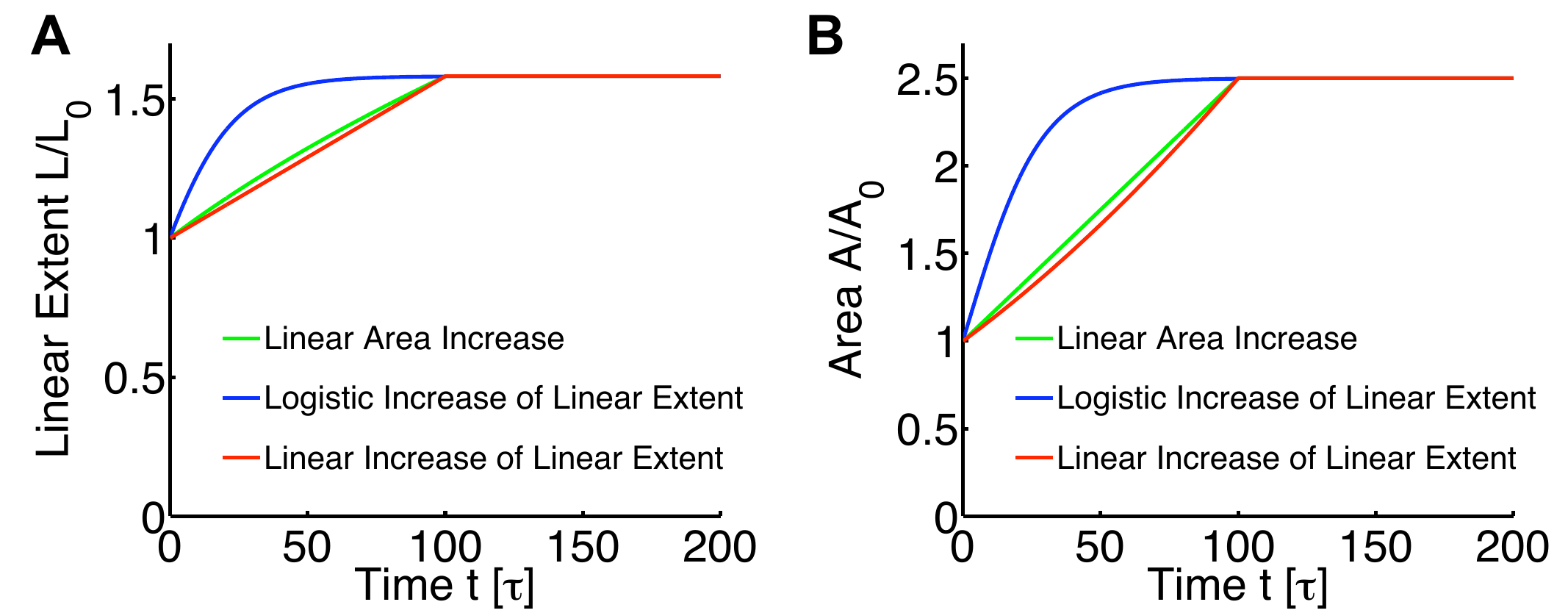} 
\caption{\textbf{Three different simulation protocols of realistic cortical
growth.} (A and B) Time courses of the linear extent $L/L_{0}$ (A) and
the area size$A/A_{0}$ (B) for linear increase of area (green traces),
linear increase of the linear extent (red traces)  and logistic growth
(blue traces). Simulations of OD development for all three growth
protocols yield qualitatively similar time courses of column spacing
$\Lambda$, number of hypercolumns $N_{HC}$ and bandedness $\alpha$.
Linear area increase and linear increase of the linear extent are
virtually identical and both provide a reasonable fit to the data
on cat V1 growth in \citep{Duffy:1998p2219} when setting 1day $\approx 1\tau$. Our quantitative characterization
of OD development in the EN model with growth (Fig. 4 in the manuscript, Fig. \ref{fig:cont_expansion_stripes}) was carried out using linear increase of the linear extent (red traces).
\label{fig:Different-Scenarios-of-growth}}
\end{figure*}

\textbf{Different simulation protocols for realistic cortical growth.}
We tested three different simulation protocols to approximate the
postnatal growth of cat V1 (see Fig. \ref{fig:Different-Scenarios-of-growth}A,
B):

\noindent (i) linear increase of the linear extent with slope $\alpha$,
starting from $L_{0}$, \[
L(t)=L_{0}+\alpha t\,,\]

\noindent (ii) sublinear increase of the linear extent $L$, such that the
simulated area $A$ increases linearly with slope $\beta$,\[
L(t)=\sqrt{L_{0}^{2}+\beta t}\,,\]

\noindent (iii) and logistic increase \citep{Crampin:1999p4188} of the linear
extent $L$\[
L(t)=L_{0}\frac{e^{\varepsilon t}}{1+\xi^{-1}(e^{\varepsilon t}-1)}\,\,\,\,\,\,\,\,\,\,\lim_{t\rightarrow\infty}L(t)=\xi L_{0}\,.\]
In each protocol, we tested total increases  between $t=0\tau$
and $t=100\tau$ by factors of 1.2, 1.4, 1.6 and 1.8 in linear extent of the simulated
region (corresponding to isotropic total area increases by a factor of 1.44, 1.96, 2.56 and 3.24). Note that according to \citep{Duffy:1998p2219},
the area of cat V1 increases between week 1 and week 12 by approximately
a factor of 2.5 and thus our range of parameters includes realistic growth conditions. 
In our simulations of OD development, we find that different growth
protocols yield similar results. In fact, the linear and sublinear
increases describe virtually identical growth scenarios 
(Fig. \ref{fig:Different-Scenarios-of-growth}, A and B) and provide
a close match to the data on V1 growth shown in \citep{Duffy:1998p2219}.
Our systematic quantitative characterization of OD development in
the EN model with growth (Fig. 4 in the manuscript) was based on the scenario of linear increase of the linear extent (red traces in Fig. \ref{fig:Different-Scenarios-of-growth}, A and B).

To model realistic growth conditions in our simulations, we linearly increased the linear extent $L$ of the simulated regions between $t=0\tau$ and $t=100\tau$. System length $L$, the Laplacian $\Delta$,
and the number of stimuli $N_{s}$ were updated at each integration
step. After $100\tau$, integration was continued for another 100$\tau$
with fixed linear extent. 

To capture and quantify the size of bandedness drops, we calculated their strength $\Delta\alpha$ and their duration $\Delta t$, i.e. the time over which $\alpha$ persistently decreased (Fig. 4E). We fitted an 8th-order polynomial to each $\alpha$-time series (least square fit). Using an 8th-order polynomial, we ensured that the fit closely followed the coarse-grained bandedness time course for all simulations.
In our simulations, we approximated the stimulus average in EN model equation (eq. (\ref{eq:dynamics_od_general_suppl})) by a representative sample of stimuli (typically between 4x10$^4$ and 2x10$^5$) drawn at each integration step. Thus, our simulations provided a stochastic approximation to the deterministic dynamics of the EN model as analyzed here. The time courses of wavelengths, bandedness and number of hypercolumns were fluctuating around the "real" time course.  Specifically, for simulations carried out for small $r $), bandedness fluctuations were around 0.01. To reliably detect growth-induced reorganizations, we only considered bandedness drops of minimum size $\Delta\alpha>0.05$. Furthermore, as growth-induced reorganization is expected to evolve on timescales several fold larger than the intrinsic time scale $\tau$ of OD segregation (see Fig. 2 and 3 in the manuscript), we only included drops with $\Delta t>15\tau$  in our analysis. The results presented in Fig. 4F-H of the manuscript were not sensitive to this particular choice of parameters.
\newline
\noindent
%
%
%
%
%
%
%
%
%
%
\section{Zigzag and Eckhaus instability  -  parameter regimes}
In this section, we discuss two fundamental instabilities of stripe patterns subject to size increase or decrease, namely the zigzag (ZZ) instability and the Eckhaus instability. We  describe their basic mechanisms and their expected parameter regime of occurrence.  As it turns out, the ZZ instability constitutes the generic behavior  upon isotropic continuous area increase.
\subsection{Zigzag instability}
The ZZ instability has been observed in many inanimate dynamical pattern forming systems as, for
instance, Rayleigh-Benard convection, when these systems experience
an instantaneous isotropic size increase \citep{manneville_90,Cross:1993p922,cross_greenside_09}. 
As pointed out in our manuscript, the result of a ZZ instability can be understood best by considering
a simple pattern of OD columns consisting of alternating stripes (Fig.
3A in the manuscript). After an isotropic area increase,
the spacing between the OD stripes is larger than the spacing set
by the Mexican-Hat. Subsequently, the structure rearranges in a sinusoidal
fashion: Through a zigzag-like bending of stripes  the original spacing
is largely recovered (see manuscript for details).

\subsection{Eckhaus instability}
The Eckhaus instability is a dynamical instability observed in systems that undergo drastic changes in system size \citep{manneville_90,Cross:1993p922,cross_greenside_09}. 
Rearrangements of domains consist
either of inserting new domains (area increase) or removing existing
ones (area decrease). Even in two-dimensional
systems, this instability gives rise to an effectively one-dimensional reorganization. It does
not require rotation symmetry of the instability mechanism. 
A recent  one-dimensional cortical growth study focused on the 
Eckhaus instability \citep{Oster:2006p1955}. However, in our simulations, we did not 
observe an Eckhaus instability, even in the case of large instantaneous 
increase or decrease of area size.
\begin{figure*}
\centering \includegraphics[width=13cm]{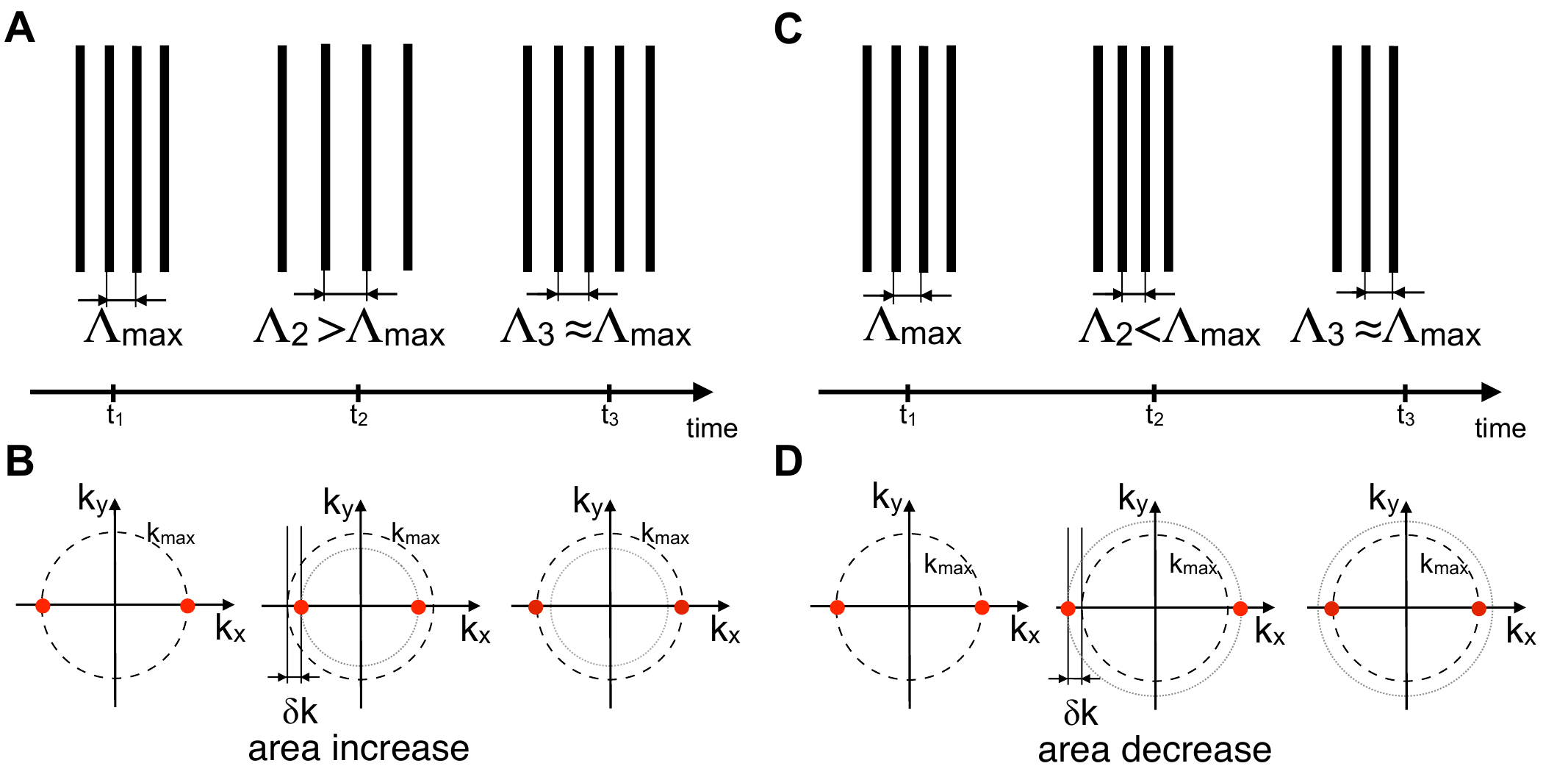} 
\caption{\textbf{Eckhaus reorganization upon large isotropic area increase
(A, B) or area decrease (C, D).} Upon area increase/decrease, OD stripes
recover their initial spacing $\Lambda_{\max}$ by insertion/removal
of OD stripes. (A) During isotropic area increase, the OD pattern
transiently acquires a spacing $\Lambda_{2}=\Lambda_{\max}+\delta\Lambda$,
with $\delta\Lambda>0$. By insertion of an new stripe the system
recovers a spacing $\Lambda_{3}$, roughly equal to $\Lambda_{\max}$.
(B) In Fourier-space, the initial pattern is centered at $\pm k_{\max}\tilde{\mathbf{x}}$.
$\delta\Lambda>0$ corresponds to a shift of $\delta k<0$. The initial
spacing is recovered by the growth of new Fourier modes close to modes
of the original pattern $\pm k_{\max}\tilde{\mathbf{x}}$. (C) During
isotropic area decrease, the OD pattern transiently acquires a spacing
$\Lambda_{2}=\Lambda_{\max}+\delta\Lambda$, with $\delta\Lambda<0$.
By removal of a stripe the system recovers a spacing $\Lambda_{3}$,
roughly equal to $\Lambda_{\max}$. (D) In Fourier-space, $\delta\Lambda<0$
corresponds to a shift of $\delta k>0$.
The initial spacing is recovered by the growth of new Fourier modes
at the center of the original pattern, i.e. $\pm k_{\max}\tilde{\mathbf{x}}$.
Note that these modes of reorganization require strong and sudden changes in area.
 \label{fig:Eckhaus-reorganization}}
\end{figure*}

The result of an Eckhaus instability can be understood best by considering
a simple pattern of OD columns consisting of alternating stripes 
(Fig. \ref{fig:Eckhaus-reorganization}). After an isotropic area
increase, the spacing between the OD stripes is larger than the spacing
set by the Mexican-Hat. Subsequently, new stripes of OD domains emerge
between existing ones by which the original spacing is eventually
recovered (Fig. \ref{fig:Eckhaus-reorganization}A).
Fig. \ref{fig:Eckhaus-reorganization}B illustrates
this in Fourier space. The initial stripe pattern is described by
a wave $\sim e^{i\tilde{\mathbf{k}}_{\max}\hat{\mathbf{x}}}$ with spatial wave vector $\tilde{\mathbf{k}}_{\max}$
and wave number $|\tilde{\mathbf{k}}_{\max}|= k_{\max}=2\pi/\Lambda_{\max}$. Upon area increase, its wavelength increases, and therefore its wave number differs by an amount $\delta k<0$ from $k_{\max}$. During an Eckhaus instability, two
Fourier modes with wave vector $\pm k_{\max}\tilde{\mathbf{x}}$ grow,
representing the new OD pattern with similar spacing as the original pattern.

Similarly, after an isotropic area decrease, the spacing between the
OD stripes is smaller than the spacing set by the Mexican-Hat. Subsequently,
a fraction of the OD stripes is removed and the original spacing is
eventually recovered (Fig. \ref{fig:Eckhaus-reorganization}C).
Fig. \ref{fig:Eckhaus-reorganization}D illustrates
this in Fourier space. Upon area decrease, the OD pattern's wave number
differs by an amount $\delta k>0$ from $k_{\max}$. Two Fourier modes
grow with wave vector $\pm k_{\max}\tilde{\mathbf{x}}$, representing
the new OD pattern in Fig. \ref{fig:Eckhaus-reorganization}C
with similar spacing as the original pattern.
%

\subsection{Parameter regimes}
In this section, we discuss the generic regions of occurrence of zigzag and Eckhaus instability in the ($\delta k,\, r)$-parameter space, where $r$ is the control parameter (or bifurcation parameter) of the system expressing the distance from the instability threshold of pattern formation, i.e. the distance from the primary finite wavelength instability, and $\delta k$ controls the size of area increase/decrease.
\begin{figure*}[htbp]
\begin{center}
\includegraphics[width=10cm]{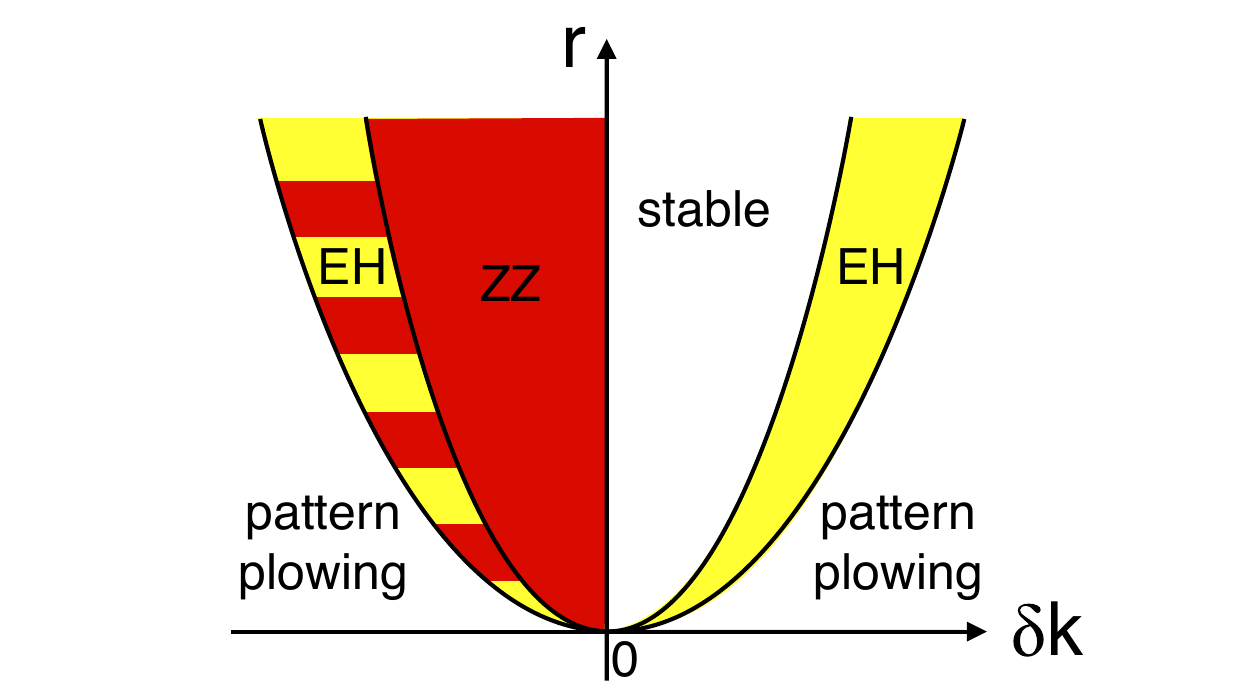}
\end{center}
\caption{{\bf Generic regions of linear instabilities (Busse Balloon) of a stripe pattern in the
($\delta k,\, r)$-plane} (redrawn from \citep{manneville_90, cross_greenside_09}). The diagram applies to the general model class of 2-dimensional relaxational, isotropic dynamics with linear Mexican-Hat type interactions.
The region of the zigzag instability (ZZ, red) is to the left of
the $\delta k=0$-axis, and is bounded by a parabola. EH (yellow)
denotes the regions of the  Eckhaus instability (see text) where a new OD stripe
is inserted ($\delta k<-\sqrt{r/3}$) or withdrawn $(\delta k>\sqrt{r/3})$.
For $-\sqrt{r}<\delta k<-\sqrt{r/3}$ these two regions overlap (striped
region). Note that the ZZ instability occurs for an arbitrarily small
area increase. Therefore, it is the generic behavior
of models for OD column formation upon continuous isotropic area increase.
\label{fig:zigzag_schematic}}
\end{figure*}
Assuming a plane wave solution $A(x,y,t)e^{ik_{\max}x}$, close to
the finite wavelength instability its behavior is governed by the Newell-Whitehead
equation \citep{newell_69} \begin{equation}
\partial_{t}A=rA+(\partial_{x}-i\partial_{y^{2}})^{2}A-|A|^{2}A\,.\label{eq:newell_whitehead_eq}\end{equation}
 The form
of Eq. \eqref{eq:newell_whitehead_eq} is independent of the
microscopic details of the instability mechanism \citep{Cross:1993p922,cross_greenside_09}.
Inserting the ansatz $A=|A|e^{i(\delta kx+\phi_{0})}$ yields a uniform
solution $|A|=\sqrt{r-\delta k^{2}}$ as long as $|\delta k|<\sqrt{r}$.
The wave number of this solution differs by an amount $\delta k$ from
$k_{\max}$, i.e. the characteristic spacing of the solution is either
smaller or larger than $\Lambda$. Linear stability analysis yields
the behavior of this solution depending on the control parameter $r$
and the difference in wave vector $\delta k$ \citep{busse_71}. Considering
general perturbations of the form $\delta A\sim e^{iq_{y}y}e^{iq_{x}x}$,
the uniform solution is found to be unstable against perturbations
with $q_{x}=0$ as soon as
\[
\delta k<0\,.
\]
This defines the domain of the ZZ instability (red regions
in Figure S\ref{fig:zigzag_schematic}). In contrast, for $q_{y}=0$
the uniform solution is only unstable if $|\delta k|>\sqrt{r/3}$.
This is the domain of the so-called Eckhaus instability (EH; yellow
regions in Figure S\ref{fig:zigzag_schematic}). For
$-\sqrt{r}<\delta k<-\sqrt{r/3}$ EH and ZZ overlap (striped region
in Fig. \ref{fig:zigzag_schematic}) and model specific
properties determine which of the two behaviors is observed in this
regime. For $|\delta k|>\sqrt{r}$, Eq. \eqref{eq:newell_whitehead_eq}
has no uniform stationary solution. In this regime, the initial stripe
pattern decays exponentially and a new OD pattern emerges  (pattern
plowing, see Fig. \ref{fig:zigzag_schematic}). 
In summary, the ZZ instability is expected to dominate the response of OD patterns if the area increase is moderate. The Eckhaus instability is expected to occur only in systems with abrupt and dramatic size changes. 
We note that the form of the diagram in Figure \ref{fig:zigzag_schematic} is very general, applying to the class of two dimensional translation and rotation symmetric relaxational dynamics in which the pattern forming process is based on a finite wavelength instability \citep{Cross:1993p922,cross_greenside_09}.
%
%
%
%
\section{ZZ reorganization during continuous isotropic area increase starting from OD stripes}
\begin{figure*}
\centering \includegraphics[width=12cm]{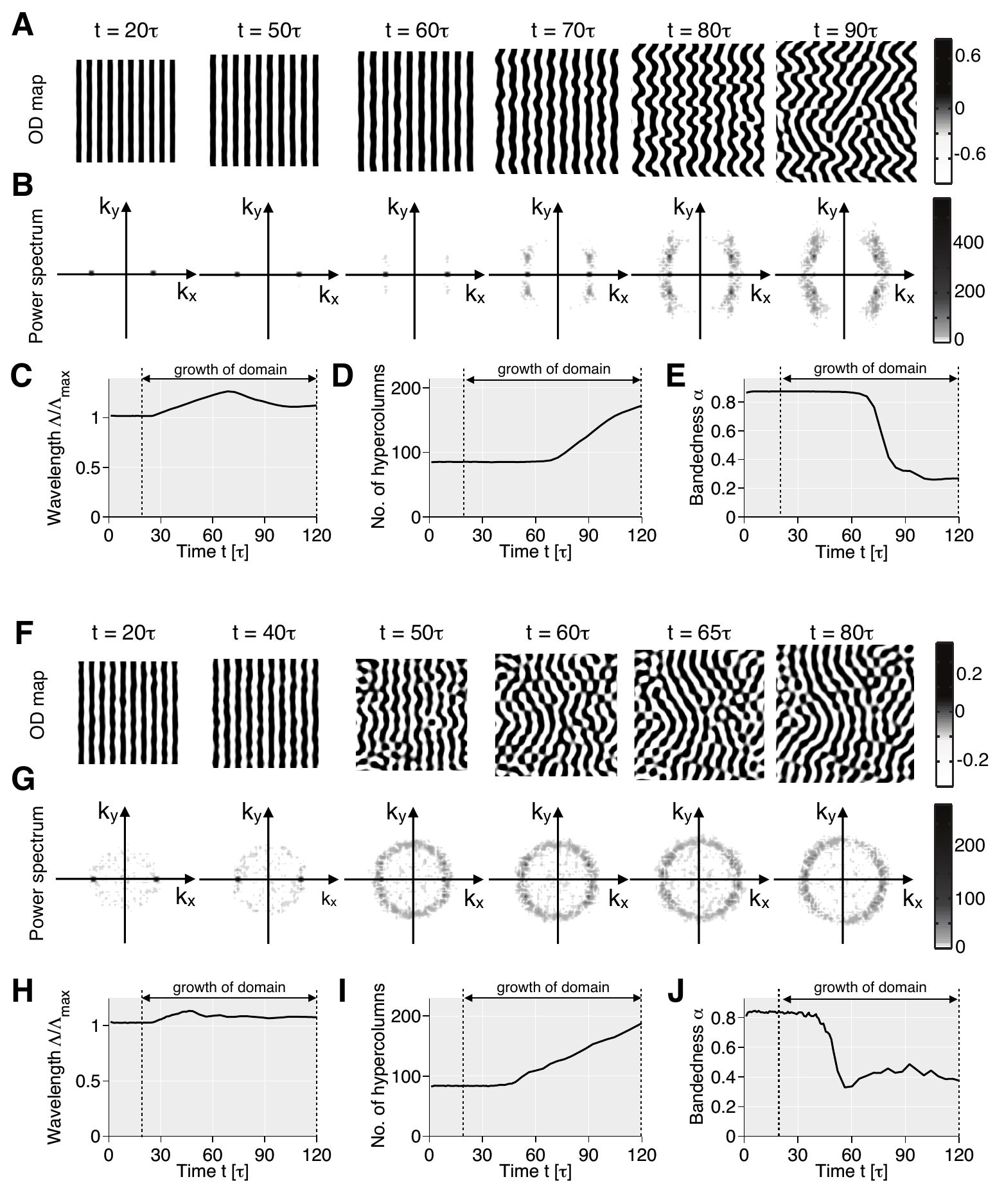}
\caption{\textbf{ZZ reorganization during continuous isotropic area increase starting from OD stripes.} 
(A and B) Snapshots of OD layouts (A)  and power spectra (B) of EN model simulations with  
continuous expansion starting from OD stripes in a parameter regime relatively far away 
from the pattern formation instability threshold  ($r=0.16$, $\eta=0.025$, total area
increase by factor of 2.56 between $t=20\tau$ and $t=120\tau$). Until $t=50\tau$
the OD stripes remain almost unchanged. Subsequently, a ZZ-type reorganization
is apparent from the sinusoidal bending of domains (A) and the growth
of Fourier Modes at $k_{\max}\tilde{\mathbf{x}}\pm q\tilde{\mathbf{y}}$
(B). ({C to E}) Time courses of column spacing
$\Lambda$ ({C}, normalized), number of hypercolumns $N_{HC}$ ({D}),
and bandedness $\alpha$ ({E}).
(F) A simulation in a regime closer to the pattern formation  threshold ($r=0.04$). 
Compared to A, the pattern acquires a more isotropic layout over time. (G) Power spectra for (F).
(H to J) Column spacing
({H}), hypercolumn number $N_{HC}$ ({I}),
and bandedness $\alpha$ ({J}).
Note, that despite a rather complex reorganization, key
signatures of a ZZ instability are visible, such as a sinusoidal bending of 
domains.\label{fig:cont_expansion_stripes}}
\end{figure*}
In our manuscript, we analyzed OD stripes subject to instantaneous 
size increase and continuous expansion starting from the nonselective cortex in the EN model. In this section, we probe the reorganization of OD stripes under continuous increase of the system size for two different values of the control parameter $r$ (Fig. \ref{fig:cont_expansion_stripes}). 
We initialized our simulations with a sinusoidal pattern $o(\mathbf{x})\sim\sin(k_{\max}x)$. After
a brief relaxation period of 20$\tau$, we linearly increased the linear
extent of the simulated regions by a factor of 1.2, 1.4, 1.6 and 1.8 between
$t=20\tau$ and $t=120\tau$ (corresponding to isotropic total area increases by a factor of 1.44, 1.96, 2.56 and 3.24). In these simulations, columnar layouts generally show reorganization on similar time scales but more irregular reorganization when compared to the scenario of instantaneous
area increases (Fig. 2, 3 in the manuscript).  
As for the case of instantaneous size increase, the column spacing $\Lambda$  
increases only transiently  (Fig. \ref{fig:cont_expansion_stripes}, C and H). Thus, the number of hypercolumns $N_{HC}$ strongly increases over the simulated time period (Fig. \ref{fig:cont_expansion_stripes}, D and I). 
The bandedness $\alpha$ decreases after onset of reorganization as expected from the
sinusoidal bending of OD domains (Fig. \ref{fig:cont_expansion_stripes}, E and J). 
Thus, a  sinusoidal bending of OD stripes which is characteristic for a ZZ instability   
also occurs upon continuous increase of the system size.
%
%
%
%
%
\section{Increasing the effective intracortical interaction width during simulations of realistic growth scenarios}
To investigate growth-induced reorganization in our manuscript, we assumed \textit{constant} EN model parameters in simulations of cortical growth. This implies that the range of lateral interaction does not change in size during cortical expansion. These interactions are of Mexican-Hat (MH) type and arise from the interplay between co-activation of cortical regions of roughly columnar size and a tendency of neighboring neurons to acquire similar response properties.

\begin{figure*}
\centering \includegraphics[width=11.5cm]{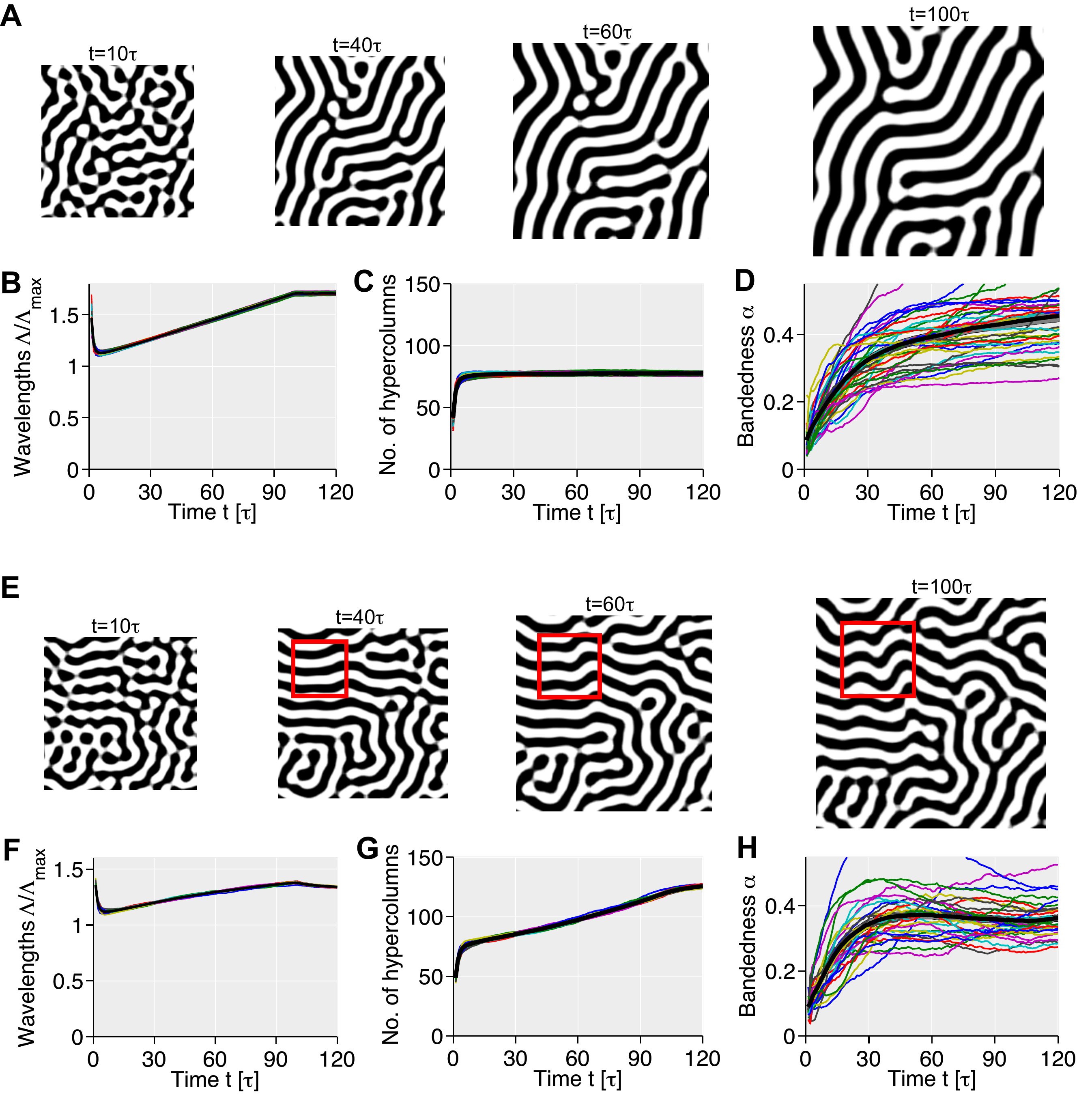}
\caption{\textbf{Increasing the width of interactions in the EN model during simulations of realistic growth scenarios.} 
(A) Snapshots of OD layouts of EN model simulations with continuous cortical expansion and continuously increasing width of the effective intracortical interactions (Mexican-hat) with the same rate ($r=0.16$, $\eta=0.025$, total area increase by a factor of 2.56 between $t=0\tau$ and $t=100\tau$).
(B to D) Time courses of column spacing $\Lambda$ (B), number of hypercolumns $N_{HC}$ (C),
and bandedness $\alpha$ (D) for N = 30 simulations (parameters as in A). The mean column spacing $\Lambda$ increases strongly between $t = 10\tau$ and $t = 100\tau$. Consequently, the number of hypercolumns $N_{HC}$ remains roughly constant.  Cortical expansion does not induce a drop in bandedness.  $\alpha$-time-courses show a monotonous increase as in simulations with non-growing cortical size (Fig. \ref{fig:elastic_net_model}). 
(E) Snapshots of OD layouts of EN model simulations with continuous expansion and continuously increasing width of the effective intracortical interactions with \textit{half} the cortical growth rate  ($r=0.16$, $\eta=0.025$, total area increase by a factor of 2.56 between $t=0\tau$ and $t=100\tau$).
(F to H) Time courses of column spacing $\Lambda$ (F), number of hypercolumns $N_{HC}$ (G),
and bandedness $\alpha$ (H) for N = 30 simulations ($r=0.16$, $\eta=0.025$, total area increase by a factor of 2.56 between $t=0\tau$ and $t=100\tau$). Growth typically induces a considerable drop in $\alpha$. However, in contrast to a fixed Mexican-Hat size (Fig. 4B in the manuscript), the mean column spacing $\Lambda$ increases considerably between $t = 10\tau$ and $t = 100\tau$. Consequently the number of hypercolumns $N_{HC}$ only increases moderately.  
\label{fig:partial_MH_scaling}}
\end{figure*}

In this section, we examine the behavior of the EN model if the interaction width, i.e. the width of the MH, is also expanding during cortical growth.
We start by considering the extreme case, where both, cortical size and interaction width increase at the same rate. Fig. \ref{fig:partial_MH_scaling}A shows the development of OD layouts for one such simulation ($r=0.16$, $\eta=0.025$, total area increase by a factor of 2.56 between $t=0\tau$ and $t=100\tau$). By mere visual inspection, an increase in column spacing can be observed. Furthermore, the OD layout rapidly becomes stripe-like during the time course. Quantifying these observations, the mean column spacing $\Lambda$ increases considerably (Fig. \ref{fig:partial_MH_scaling}B), the number of hypercolumns $N_{HC}$ is roughly constant (Fig. \ref{fig:partial_MH_scaling}C) and the bandedness $\alpha$ increases persistently (Fig. \ref{fig:partial_MH_scaling}D) as in the simulations without growth (Fig. \ref{fig:elastic_net_model}F).

What happens if the interaction width increases at lower rate than the cortex?  Fig. \ref{fig:partial_MH_scaling}E shows the typical development of OD layouts in a simulation for which MH growth rate was set to half of the cortical area growth rate ($r=0.16$, $\eta=0.025$, total area increase by a factor of 2.56 between $t=0\tau$ and $t=100\tau$).
In contrast to Fig. \ref{fig:partial_MH_scaling}A, the OD layout acquires a more bended and ZZ-type shape during the time course, strikingly similar to Fig. 4A in the manuscript (see red frames in Fig. \ref{fig:partial_MH_scaling}E).
The mean column spacing $\Lambda$ increases considerably (Fig. \ref{fig:partial_MH_scaling}F), though less than in Fig. \ref{fig:partial_MH_scaling}B. The number of hypercolumns
$N_{HC}$ increases (Fig. \ref{fig:partial_MH_scaling}G), however less than in Fig. 4C of the manuscript. The bandednesses $\alpha$ exhibits a pronounced decrease, similar to the growing systems simulated in Fig. 4 - a characteristic feature of the ZZ instability.

The above observations are consistent with our results on simulations with fixed interaction width (see manuscript). There we have shown that even moderate cortical expansion rates can induce substantial cortical reorganization, both in simulations with abrupt size increases as well as in realistic growth scenarios. However, in the case at hand, this is accompanied by an increase in mean column spacing, and a relatively small increase in the number of hypercolumns. As we did not observe any increase in column spacing in cat V1 between week 4 and 14, this suggests that the effective intracortical interaction width stays approximately constant (see discussion in the manuscript).

\bibliographystyle{unsrt}

\end{document}